\documentclass[preprint,12pt]{elsarticle}

\usepackage{amssymb}
\usepackage{amsmath}
\usepackage{graphicx}

\usepackage{picinpar}
\usepackage{amsfonts}
\usepackage{url}
\usepackage{flushend}
\usepackage[latin1]{inputenc}
\usepackage{colortbl}
\usepackage{soul}
\usepackage{multirow}
\usepackage{pifont}
\usepackage{color}
\usepackage{alltt}
\usepackage[hidelinks]{hyperref}
\usepackage{enumerate}
\usepackage{siunitx}
\usepackage{breakurl}
\usepackage{epstopdf}
\usepackage{pbox}
\usepackage{subfig}
\usepackage{caption}
\usepackage[linesnumbered,ruled,vlined]{algorithm2e}

\journal{Advanced Engineering Informatics}

\begin{document}

\begin{frontmatter}



\title{Physics-Informed Neural Network for Cross-Domain Predictive Control of Tapered Amplifier Thermal Stabilization}

\author{Yanpei Shi\fnref{label1}}
\author{Bo Feng\fnref{label2}}
\author{Yuxin Zhong\fnref{label1}}
\author{Haochen Guo\fnref{label1}}
\author{Bangcheng Han\fnref{label1}}
\author{Rui Feng\corref{cor1}\fnref{label1}}

\ead{fengrui_buaa@buaa.edu.cn}
\cortext[cor1]{Corresponding author.}

\affiliation[label1]{organization={School of Instrumentation and Optoelectronic Engineering, Beihang University},
             postcode={Beijing 100191},
             country={China}}

\affiliation[label2]{organization={Fu foundation school of engineering and applied science, Columbia University in the City of New York},
             postcode={NY 10027},
             country={USA}}

\begin{abstract}
Thermally induced laser noise poses a critical limitation to the sensitivity of quantum sensor arrays employing ultra-stable amplified lasers, primarily stemming from nonlinear gain-temperature coupling effects in tapered amplifiers (TAs). To address this challenge, we present a robust intelligent control strategy that synergistically integrates an encoder-decoder physics-informed gated recurrent unit (PI-GRU) network with a model predictive control (MPC) framework. Our methodology incorporates physical soft constraints into the neural network architecture, yielding a predictive model with enhanced physical consistency that demonstrates robust extrapolation capabilities beyond the training data distribution. Leveraging the PI-GRU model's accurate multi-step predictive performance, we implement a hierarchical parallel MPC architecture capable of real-time thermal instability compensation. This hybrid approach achieves cross-domain consistent thermal stabilization in TAs under diverse laser power operations. Remarkably, while trained exclusively on low-power operational data, our system demonstrates exceptional generalization, improving prediction accuracy by 58.2\% and temperature stability by 69.1\% in previously unseen high-power operating regimes, as experimentally validated. The novel synchronization of physics-informed neural networks with advanced MPC frameworks presented in this work establishes a groundbreaking paradigm for addressing robustness challenges in cross-domain predictive control applications, overcoming conventional modeling limitations.
\end{abstract}



\begin{keyword}
Physics-informed neural network \sep gated recurrent unit \sep nonlinear model predictive control \sep semiconductor laser temperature control.


\end{keyword}

\end{frontmatter}



\section{Introduction}
\label{sec1}
Ultra-stable lasers provide quantum sensing systems with unparalleled detection sensitivity in noisy environments, thereby delivering transformative potential across medical imaging, geological exploration, and precision navigation \cite{liang2015,castro2025}. In the emerging field of biomagnetic imaging, these lasers serve as essential pumping sources for atomic magnetometer arrays employing spin-exchange relaxation-free (SERF) mechanisms, where laser spectral purity governs fidelity of weak biological signals detection \cite{brookes2022,zhang2025a}. In response to the dual requirements of spectral stability and power scaling, an optical amplification scheme using tapered amplifier (TA) is implemented \cite{faugeron2015,franken2025}. Watt-level laser supports simultaneous operation across hundreds of sensors, facilitating high-resolution reconstruction of biomagnetic signatures. Enhanced by stable amplified lasers, the multi-channel magnetometer arrays advance diagnostic capabilities for early screening of neurological and cardiovascular disorders \cite{rea2022,brickwedde2024}.

However, due to the inherent gain-temperature coupling in TAs, the thermally induced optical noise substantially degrades the low-frequency sensitivity of magnetometers crucial for weak biomagnetic fields detection \cite{kittlaus2025}. The thermal instability stems from the temperature-dependent semiconductor active region of TAs \cite{spreemann2009,wu2023a}. Resulting laser power fluctuations disrupt coherent manipulation of SERF atomic ensembles, compromising the accuracy of magnet field reconstruction \cite{long2023}. Therefore, active thermal stabilization of TAs constitutes a critical prerequisite for sustaining diagnostic-grade reliability in quantum biomagnetic sensing systems.

This work falls within the realm of semiconductor laser temperature control, with emphasis on addressing the challenges in TA laser systems. Through coordinated heating and cooling cycles, thermoelectric cooler (TEC)-based thermal regulation offers a compact and cost-effective solution for achieving millikelvin-level stability in milliwatt-level laser diodes for quantum prototype applications \cite{vicarini2019,pokryshkin2020,peng2021}. Existing control strategies predominantly employ proportional integral derivative (PID)-based approaches, with enhancements including fuzzy logic for adaptive tuning, dual-loop synergistic architectures, and analog compensations via specialized chips \cite{peng2021,xie2019,xu2020,huang2024}. Based on mechanistic models, system identification leverages polynomial fitting for transfer function derivation, combining with experimental methods such as M-sequence excitation and differential evolution algorithms \cite{wu2023a,xu2020,zhao2022a}. However, the efficacy of conventional approaches for low-power lasers considerably diminishes in high-power TA configurations, as manifested in two fundamental limitations \cite{kittlaus2025,pokryshkin2020,peng2021,akbar2013}: 
\begin{enumerate}[(1)]
	\item Time-dependent thermal hysteresis from delayed heat transfer in the TEC-based thermal management structure; 
	\item Systematic mismatches between simplified dynamic models and the observed behavior of high-power lasers across diverse operating regimes. 
\end{enumerate}
Both limitations are further exacerbated by intrinsic thermo-optic nonlinearities governing optical amplification processes, where the effects of bandgap narrowing, carrier mobility degradation, and modal gain attenuation collectively establish intricate multi-physics coupling \cite{franken2025,akbar2013,gong2016}. Consequently, existing control frameworks prove insufficient to ensure laser thermal stability required for high-power operation, primarily due to their inability to tackle the interdependent dynamics of TA systems.

Model predictive control (MPC) emerges as a promising approach, combining explicit constraint handling with inherent multi-step foresight to proactively address system inertia \cite{yan2012,guo2018,zhou2024}. Given that the effectiveness of MPC method hinges on model accuracy, developing advanced modeling techniques becomes a central research priority for nonlinear system control \cite{wang2016,fei2022,stiti2024,liang2024,sun2022}. Rising as a powerful tool for characterizing nonlinear dynamics, gated recurrent unit (GRU)-based neural networks excel at capturing temporal dependencies via streamlined gating mechanisms, while mitigating vanishing or exploding gradients in traditional recurrent neural  architectures \cite{hu2020,yuan2022,li2022,chen2023}. Furthermore, by incorporating physics-informed neural networks (PINNs), predictive models leverage physically constrained learning to achieve reliable predictions in data-sparse scenarios \cite{zhou2024,wang2024,son2023,xu2024,li2025}. However, it remains unclear how to embed PINN framework into MPC strategy for TA thermal stabilization across multi-power laser operations, which motivates this work. Key technical challenges center on two interrelated fronts: (1) designing a predictive model with high-accuracy multi-step forecasting and cross-domain generalizability; (2) developing a control framework for real-time integration of neural network inference in optimization processes. 

Main contributions of this study are summarized as follows:
\begin{itemize}[-]
	\item A physics-informed GRU (PI-GRU) network with an encoder-decoder architecture is developed in Section \ref{sec3a}, integrating thermodynamics principles via soft constraints derived from lumped-parameter thermal models. Under unseen high-power conditions, the trained PI-GRU achieves an over $58.2\%$ reduction in mean absolute error (MAE) for multi-step temperature predictions with improved single-step accuracy, significantly outperforming the data-driven counterpart, as detailed in Section \ref{sec4b}.
	\item A hierarchical parallel MPC architecture is proposed in Section \ref{sec3c} to execute multi-stage computational tasks through tiered batch processing, guaranteeing solution latency below $0.1 \text{ s}$ under real-time control constraints. At the out-of-domain operation of $1.5 \text{ W}$, PI-GRU MPC suppresses the range of 1000-s TA temperature fluctuations within $\SI{0.008}{\celsius}$, matching the thermal stability performance at the baseline scenario of $0.5 \text{ W}$, as detailed in Section \ref{sec4c}.
	\item This work establishes the first demonstration of an end-to-end deep learning framework in semiconductor laser temperature control, achieving high accuracy and cross-domain generalizability. The co-design of PINN and MPC provides a methodological paradigm for the robustness enhancement in nonlinear control systems.
\end{itemize}

The paper is organized as follows. Section \ref{sec2} analyzes core challenges in TA temperature control following system modeling. Section \ref{sec3} develops the PIGRU network and its integration with the hierarchical parallel MPC framework. Section \ref{sec4} presents comprehensive experimental validation of the proposed approach on a physical TA laser system. Conclusion and prospect are provided in Section \ref{sec5}.

\section{Problem Formulation and Outline of the Proposed Strategy}
\label{sec2}
\begin{figure}[t]
	\centering
	\subfloat[]
	{		
		\begin{minipage}[h]{\columnwidth}
			\centering
			\includegraphics[width=0.9\textwidth]{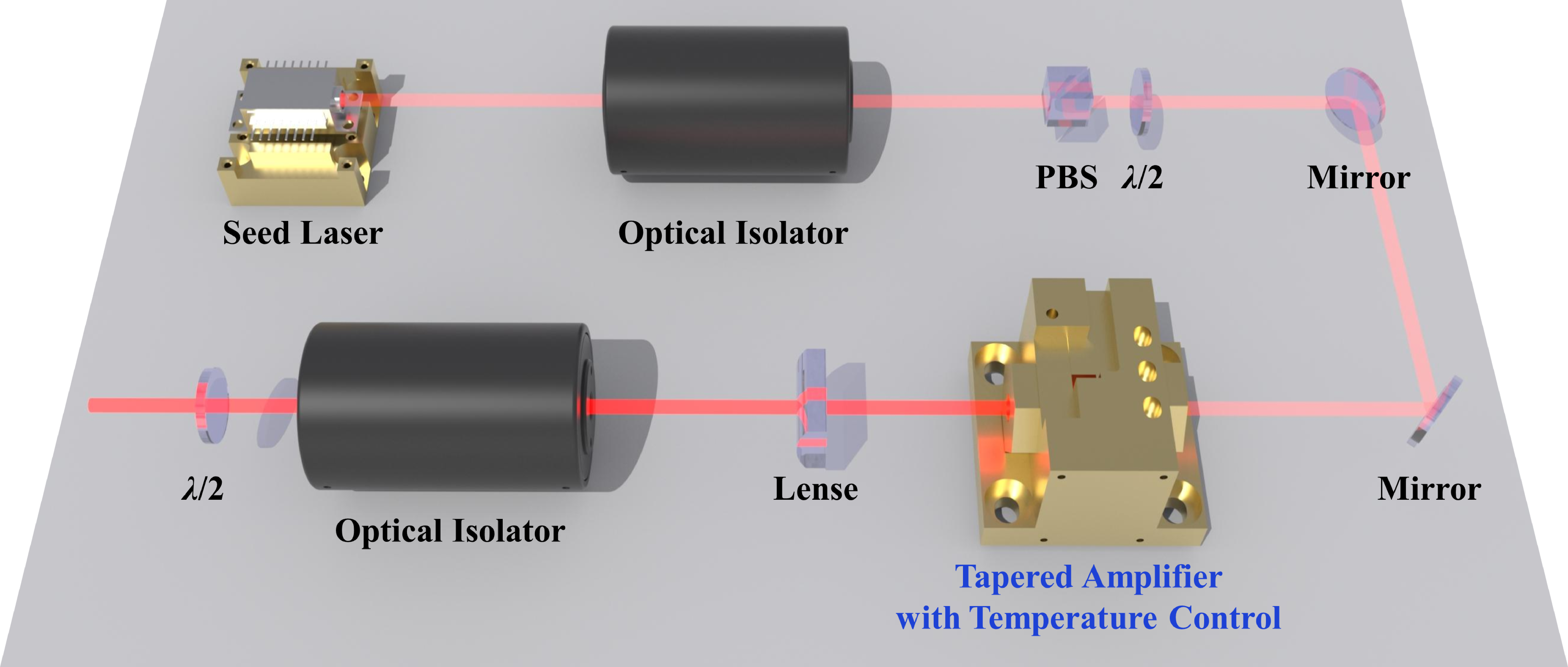}
			\label{fig1a}
		\end{minipage}
	}
	\\
	\subfloat[]
	{
		\begin{minipage}[t]{0.3\columnwidth}
			\centering
			\includegraphics[height=0.16\textheight]{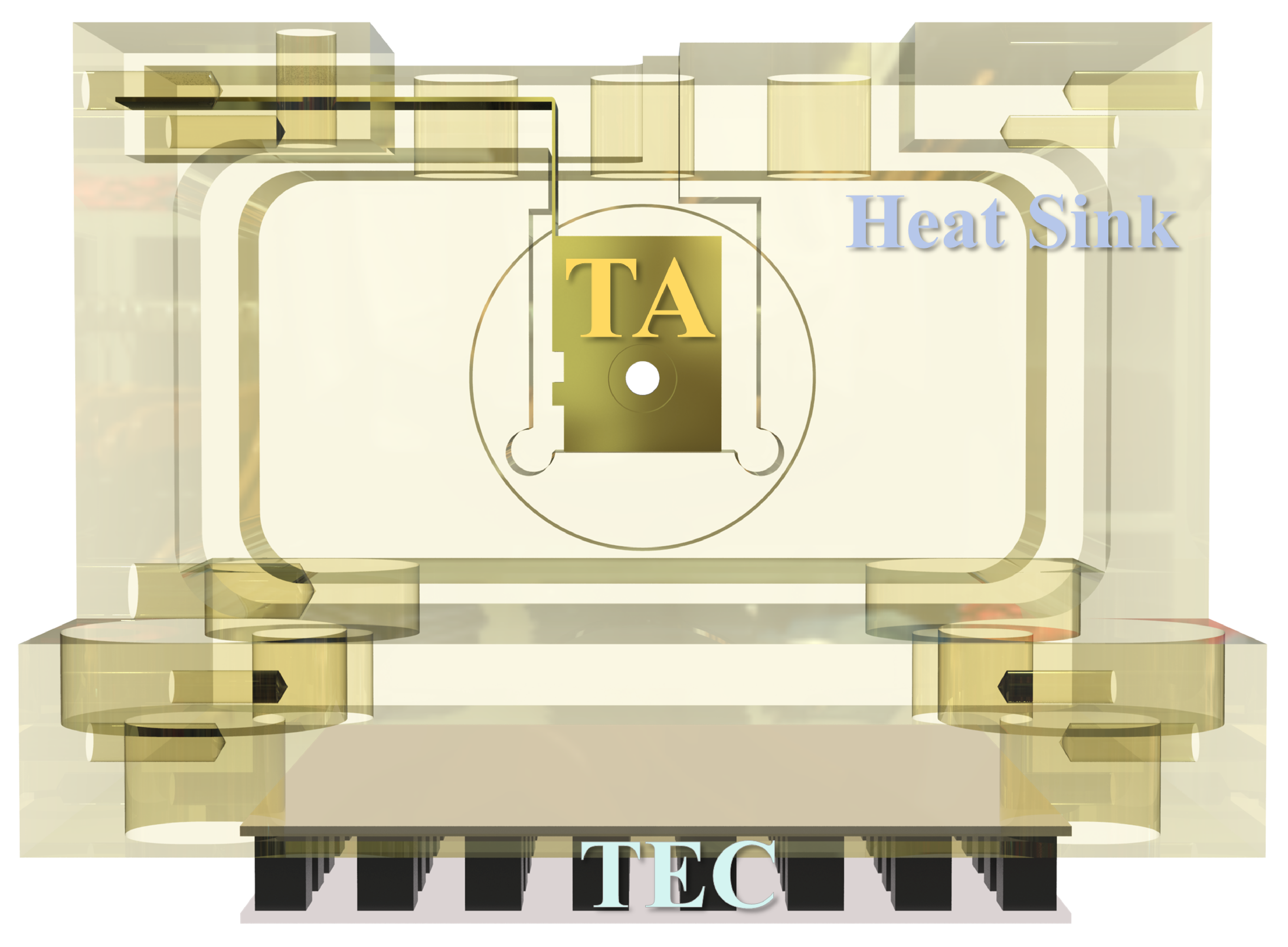}
			\label{fig1b}
			\vspace{-5mm}
		\end{minipage}
	}
	\subfloat[]
	{
		\begin{minipage}[t]{0.3\columnwidth}
			\centering
			\includegraphics[height=0.16\textheight]{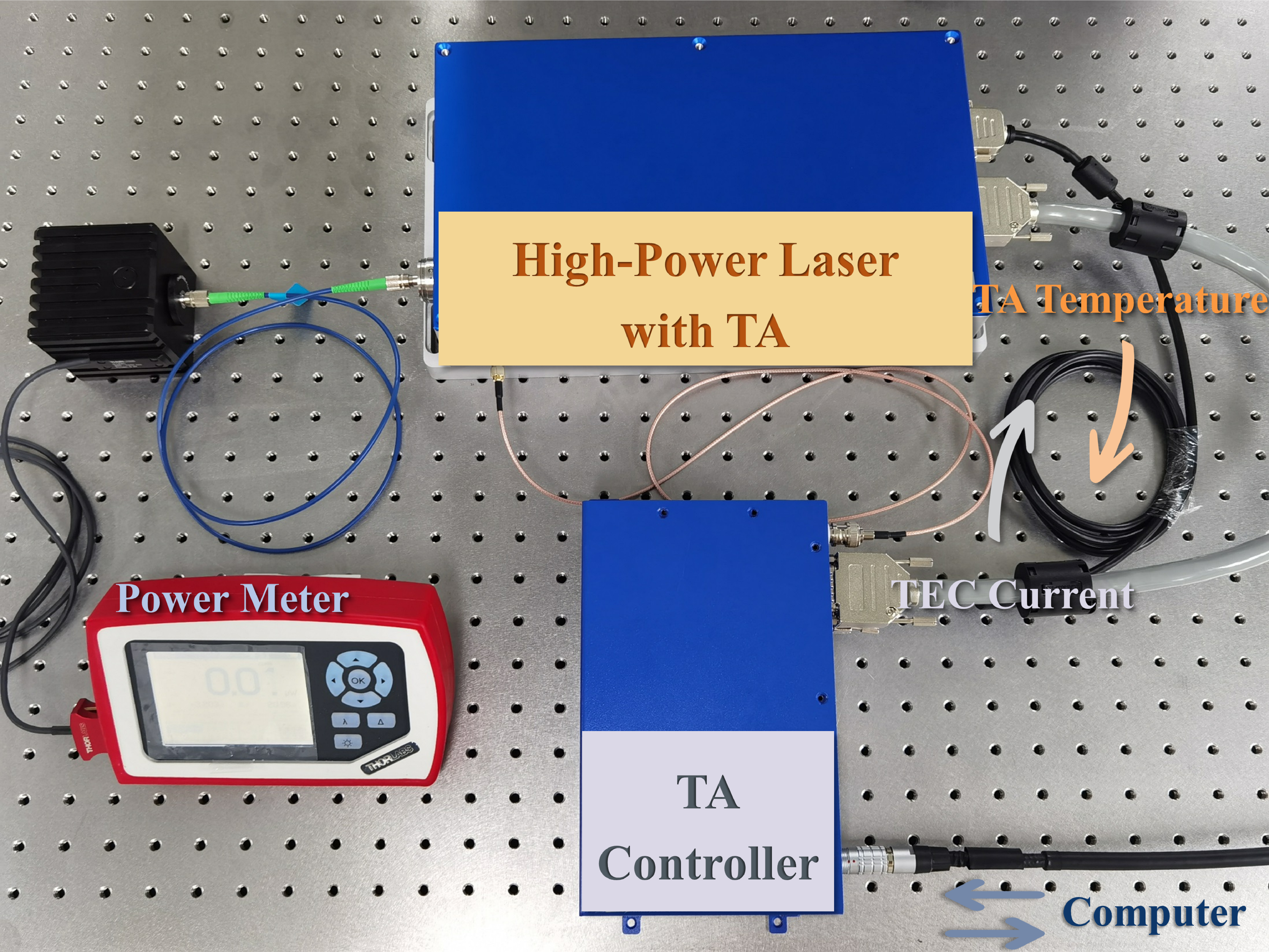}
			\label{fig1c}
			\vspace{-5mm}
		\end{minipage}
	}
	\subfloat[]
	{
		\begin{minipage}[t]{0.3\columnwidth}
			\centering
			\includegraphics[height=0.16\textheight]{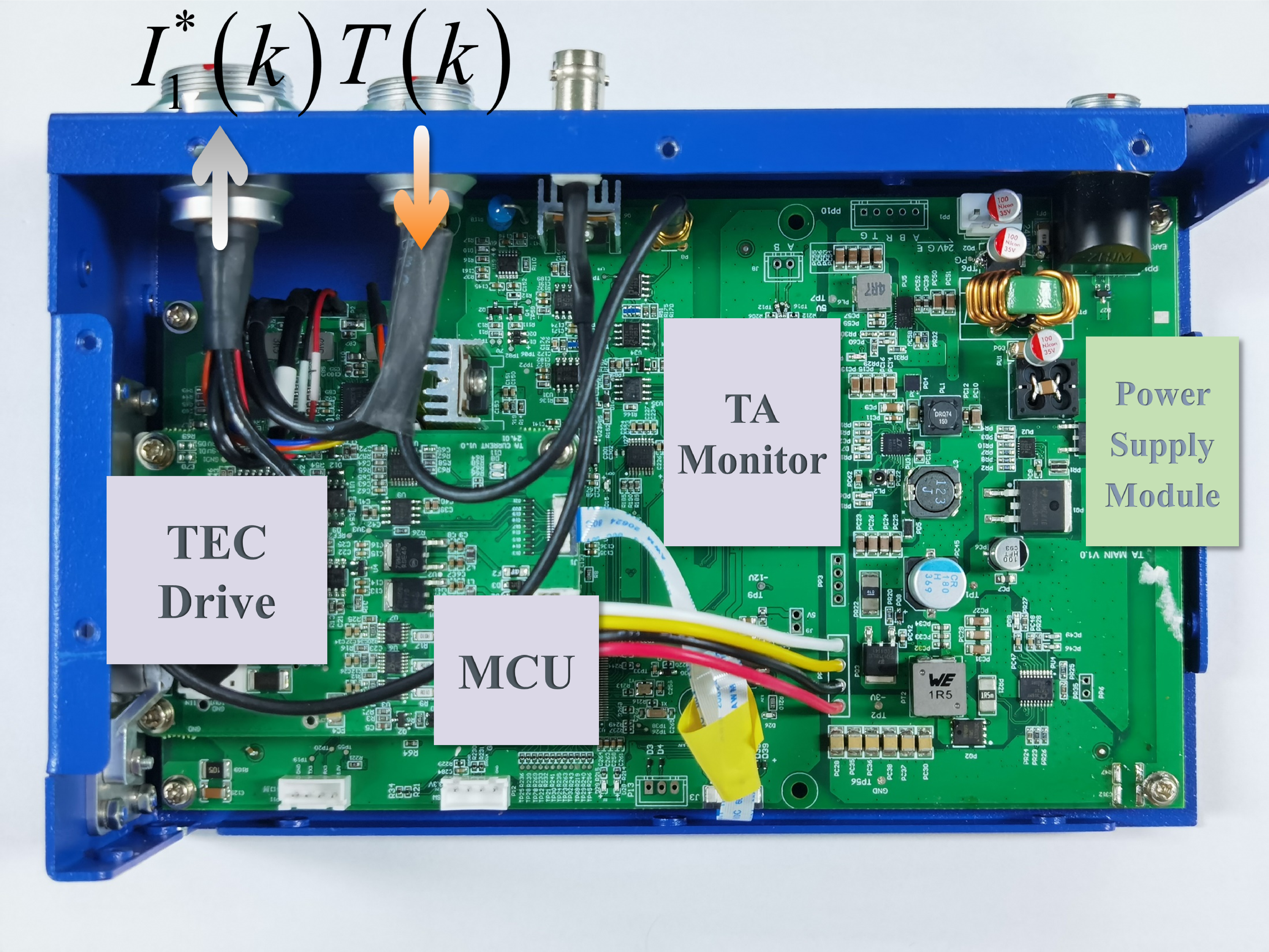}
			\label{fig1d}
			\vspace{-5mm}
		\end{minipage}
	}
	\caption{(a) Schematic of the TA-based high-power semiconductor laser system. PBS: polarizing beam splitter, ${\lambda  \mathord{\left/ {\vphantom {\lambda  2}} \right. \kern-\nulldelimiterspace} 2}$: half-wave plate. (b) Thermal management of the TA. (c) Experimental platform for TA temperature control. (d) Internal photograph of the TA controller. MCU: microcontroller unit.}
	\label{fig1}
\end{figure}

Fig. \ref{fig1}\subref{fig1a} depicts the schematic of a TA-based high-power semiconductor laser system. A single-mode seed laser undergoes amplification by a temperature-controlled TA module along the optical path. As specifically shown in Fig. \ref{fig1}\subref{fig1b}, the C-mount packaged TA chip (Toptica-eagleyard, EYP-TPA-0795-02000-4006-CMT04-0000) is integrated into a custom thermal management to ensure temperature uniformity. Here, the interface of the TEC(Laird Thermal Systems, CP08-127-06-L1-RT-W4.5) with the TA establishes thermal equilibrium via conductive coupling. Meanwhile, temperature is monitored in real time by an embedded thermistor. With directional arrows indicating signal flow, the closed-loop control is implemented on the experimental validation platform in Fig. \ref{fig1}\subref{fig1c} through the TEC current modulation and TA temperature feedback. Fig. \ref{fig1}\subref{fig1d} details the hardware realization of the synchronized data acquisition and control terminal.

We propose a neural network-enhanced MPC strategy to eliminate TA temperature fluctuations. The control strategy leverages multi-step predictions from the trained model to iteratively compute the optimal TEC current sequence ${I^*}\left( k \right) = \left( {I_1^*\left( k \right),{\text{ }}I_2^*\left( k \right),{\text{ }} \cdots {\text{, }}I_m^*\left( k \right)} \right)$, where $m$ denotes the prediction horizon length. At each sampling instant $k$, the first component $I_1^*\left( k \right)$ is applied to the TEC control input through a receding horizon implementation. The constrained optimization process is formulated as:
\begin{equation} \label{eq.1}
	I_1^*\left( k \right) = \mathcal{D}\left({T\left( k \right),{T_{{\text{ref}}}},\mathcal{G},\Xi } \right)	
\end{equation}
where $T\left( k \right)$ denotes the real-time temperature measurement, ${T_{\text{ref}}}$ represents the temperature setpoint, $\mathcal{G}$ defines the predictive model, and $\Xi$ specifies the constraint set. The inference of the $m$-step temperature prediction sequence $\hat T\left( k \right) = \left( {{{\hat T}_1}\left( k \right),{\text{ }}{{\hat T}_2}\left( k \right),{\text{ }} \cdots {\text{, }}{{\hat T}_m}\left( k \right)} \right)$ is expressed as:
\begin{equation} \label{eq.2}
	\hat T\left( k \right) = \mathcal{G}\left( {\bar T\left( k \right),\bar I\left( k \right);\theta } \right)
\end{equation}
where $\bar T\left( k \right) = \left( {{T_1}\left( k \right),{\text{ }}{T_2}\left( k \right),{\text{ }} \cdots ,{\text{ }}{T_n}\left( k \right)} \right)$ represents the $n$-step historical TA temperature profile, $\bar I\left( k \right) = \left( {{I_1}\left( k \right),{\text{ }}{I_2}\left( k \right),{\text{ }} \cdots {\text{, }}{I_m}\left( k \right)} \right)$ constitutes $m$-step TEC control action candidates, and $\theta$ denotes the trainable parameter set.  

Obtaining comprehensive data of TAs poses critical challenges, particularly in achieving consistent coverage across the entire power operating envelope. Given the risk of premature aging or irreversible optical degradation to TA chips, aggressive operational regimes under high power levels fundamentally violates laser safety protocols. This experimental data paucity necessitates developing a specialized predictive model trained exclusively on low-power samples, with generalizability across diverse operating conditions. Thereupon, based on the mentioned challenges of cross-domain TA thermal stabilization, this work aims to address the following questions:
\begin{enumerate}[(1)]
	\item How to design a PINN efficiently learning generalizable thermal dynamics representations from limited low-power training data?
	
	\item How to formulate a real-time MPC strategy achieving an optimal trade-off between control performance and computational complexity?
\end{enumerate}

\section{Physics-Informed Neural Networks in Nonlinear Model Predictive Control}\label{sec3}
\begin{figure}[!htb]
	\centering
	\includegraphics[width=0.98\columnwidth]{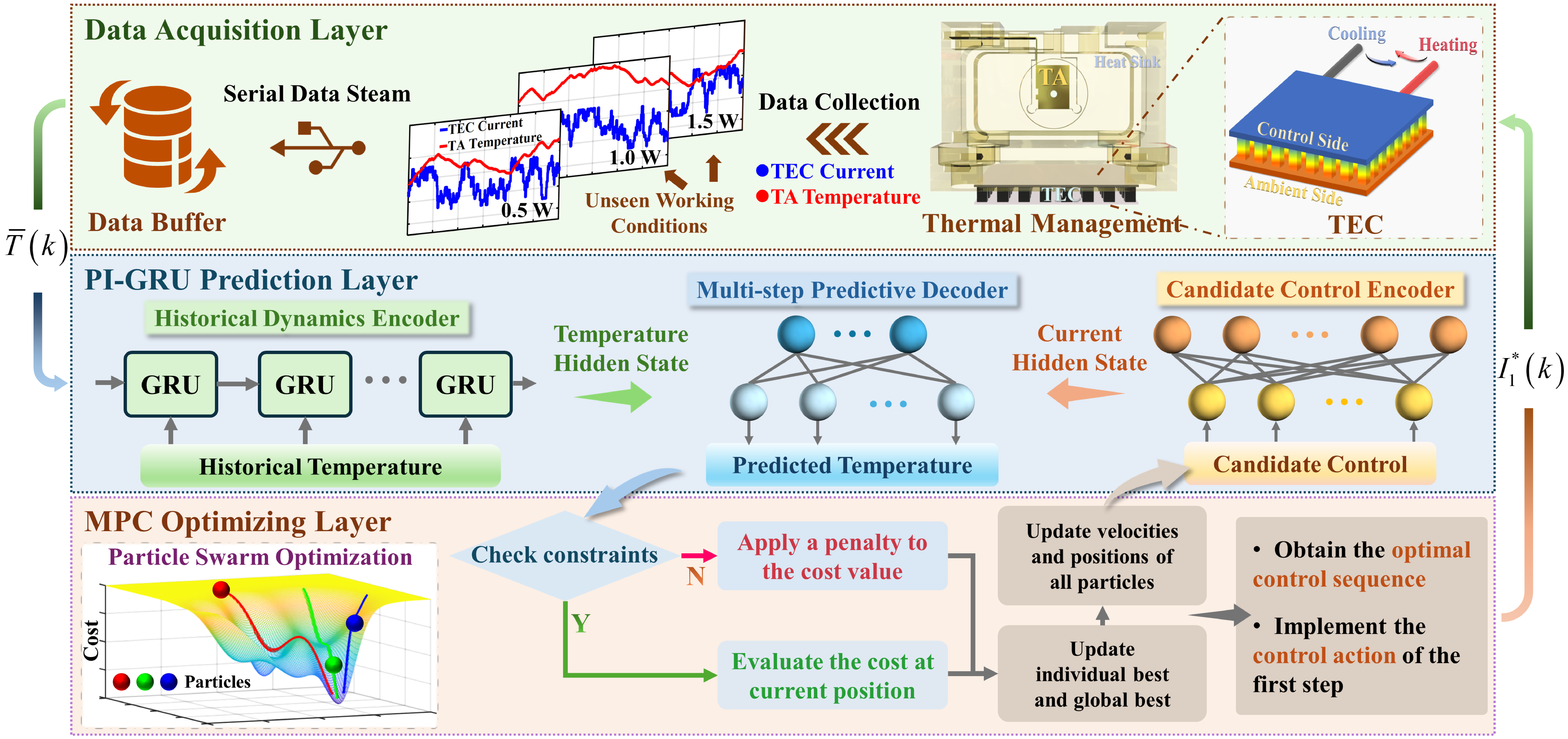}
	\caption{Framework of the proposed control strategy based on network predictive model}
	\label{fig2}
\end{figure}
In this section, we propose a specialized PI-GRU network with an encoder-decoder architecture for thermodynamics learning. The obtained temperature sequence predictions guide TEC inputs optimization, thereby facilitating closed-loop control. The implementation framework is shown in Fig. \ref{fig2}.
\subsection{Construction of PI-GRU Network}
\label{sec3a}
\begin{figure}[!tb]
	\centering
	\includegraphics[height=0.67\textheight]{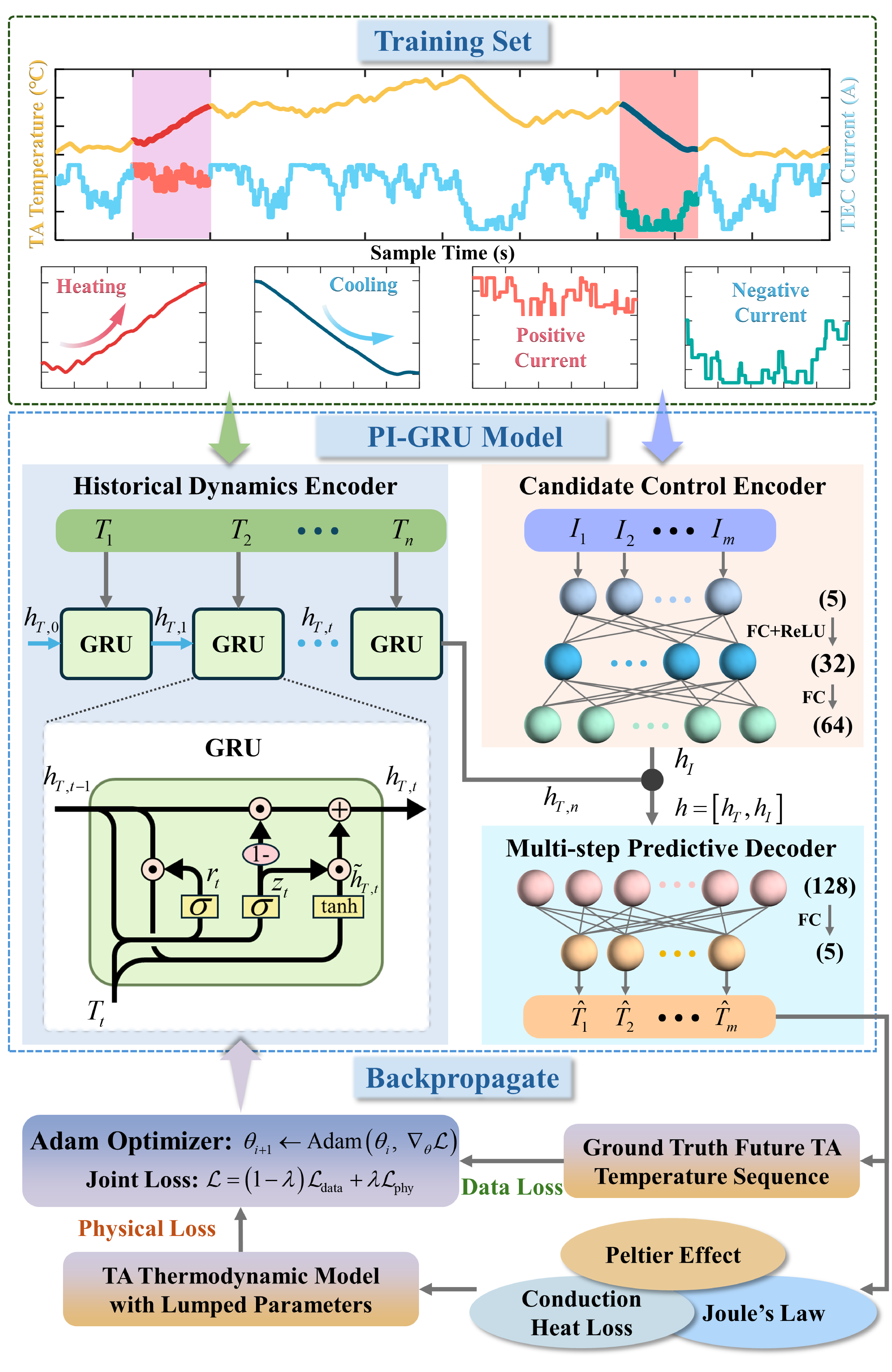}
	\caption{Construction and training process of PI-GRU network}
	\label{fig3}
\end{figure}

The PI-GRU network contains three core components: a historical dynamics encoder analyzing temporal patterns $\bar T$, a candidate control encoder condensing control inputs $\bar I$, and a multi-step predictive decoder integrating features to generate temperature sequence predictions $\hat T$, as shown in Fig. \ref{fig3}.  

\subsubsection{Historical Dynamics Encoder}
This encoder employs an $n$-dimensional GRU network specializing in temporal dependency extraction from the TA temperature sequences $\bar T$. For batch sample \(j\), hidden state \(h_{T,n}^j\) aggregates the \(n\) preceding time steps' information. The activation mechanism implements gated combination between the previous activation $h_{T,t-1}^j$ and the candidate activation \(\tilde h_{T,t}^j\):
\begin{equation} \label{eq.3}
	h_{T,t}^j = (1 - z_t^j) h_{T,t-1}^j + z_t^j \tilde h_{T,t}^j
\end{equation} 
where the update gate \(z_t^j\) regulates the balance between historical retention and new input incorporation. The fusion of existing memory and updated states follows: 
\begin{equation} \label{eq.4}
	z_t^j = \sigma {\left( {{W_z}{{\bar T}_t^j} + {U_z}{h_{T,t - 1}^j}} \right)}
\end{equation} 
where \(\sigma\) denotes the sigmoid activation, with \(W_z\) and \(U_z\) being update gate weight matrices. The candidate activation \(\tilde h_{T,t}^j\) in standard GRU architecture follows:
\begin{equation} \label{eq.5}
	\tilde h_{T,t}^j = \tanh {\left( {W{{\bar T}_t^j} + U\left( {{r_t^j} \odot {h_{T,t - 1}^j}} \right)} \right)}
\end{equation}  
where \(W\) and \(U\) are learnable weights and \(r_t\) denotes the reset gate, with \(\odot\) representing element-wise multiplication. The reset gate $r_t^j$ modulates historical memory filtering, expressed similarly to the update gate:  
\begin{equation} \label{eq.6}
	r_t^j = \sigma {\left( {{W_r}{{\bar T}_t^j} + {U_r}{h_{T,t - 1}^j}} \right)}
\end{equation}  
where \(W_r\) and \(U_r\) are the reset gate weight matrices. Modeling global dependencies, the encoder aims to learn the intrinsic dynamics of TEC-controlled TA temperature.
\subsubsection{Candidate Control Encoder}
Through dual fully connected layers, this encoder projects the TEC control inputs $\bar I$ into a hidden state \(h_I^j\) via: 
\begin{align} 
	\label{eq.7}
	h_{I_1}^j &= \text{ReLU}\left( W_{I_1}\bar I^j + b_{I_1}\right)\\
	\label{eq.8}
	h_I^j &= W_{I_2}h_{I_1}^j + b_{I_2}
\end{align} 
where \(W_{I_1}\), \(W_{I_2}\), \(b_{I_1}\), and \(b_{I_2}\) are the weights and biases of the encoder. The $\text{ReLU}$ activation facilitates MPC-oriented nonlinear dynamic feature extraction of TEC, capturing bidirectional thermal regulation through heating and cooling phase transitions. Through multi-scale fusion, the two-stage encoding architecture establishes dimensional matching between control features and temporal patterns via latent space alignment, enabling coordinated state updates in the predictive decoder.

\subsubsection{Multi-Step Predictive Decoder}
For the output layer of the PI-GRU network, a consolidated hidden vector \({h^j}\) is formed by concatenating historical thermal states $h_{T,n}^j$ and projected control embeddings $h_I^j$. The multi-step predictive decoder, with weights ${W_y}$ and bias ${b_y}$, transforms \({h^j}\) into the output temperature predictions:  
\begin{equation} \label{eq.9}
	\hat T_t^j = {W_y}h^j + {b_y} .  
\end{equation} 

\subsection{Training of PI-GRU Network}

\SetKwFor{Parallel}{parallel}{do}{end for}
\begin{algorithm}[!ht]
	\caption{Training Process of PI-GRU Model}
	\label{alg.1}
	\KwIn{Historical TA temperature sequence $\bar T$, candidate TEC current sequence  $\bar I$, ground truth future TA temperature sequence $\overset{\lower0.5em\hbox{$\smash{\scriptscriptstyle\frown}$}}{T}$, number of epochs $E$, sample size $N$, batch size $B$, loss weight $\lambda$, learning rate $\eta$,}
	\KwOut{Model parameter set $\theta$.}
	
	\LinesNumbered
	Initialization: Hidden state ${h_T} \leftarrow 0$.
	
	\For{$e = 1,{\text{ }}2,{\text{ }} \cdots ,{\text{ }}E$}
	{
		\For{$i = 1,{\text{ }}2,{\text{ }} \cdots ,{\text{ }}{N \mathord{\left/{\vphantom {N B}} \right.\kern-\nulldelimiterspace} B}$}
		{
			Sample batch of $\left( {{{\bar T}^i},{{\bar I}^i},{{\hat T}^i}} \right)$ from measured datapoints\;
			\Parallel{$j = 1,{\text{ }}2,{\text{ }} \cdots ,{\text{ }}B$}
			{
				\For{$t = 1,{\text{ }}2,{\text{ }} \cdots ,{\text{ }}n$}
				{
					Forward $T_t^{i,j}$  and update hidden state $h_{T,t}^{i,j}$ using \eqref{eq.3}--\eqref{eq.6}\;
				}
				Encode ${\bar I^{i,j}}$ to compute feature $h_I^{i,j}$ using \eqref{eq.7} and \eqref{eq.8}\;
				Concatenate features $\left( {h_{T,n}^{i,j},{\text{ }}h_I^{i,j}} \right)$ as ${h^{i,j}}$\;
				Decode ${h^{i,j}}$ to compute solution ${\hat T^{i,j}}$ using \eqref{eq.9}\;
			}
			Compute total loss $\mathcal{L}\left( \theta  \right)$ using \eqref{eq.10}--\eqref{eq.16}\;
			Backpropagate and update model parameters ${\theta _{i + 1}} \leftarrow \operatorname{Adam}\left({\theta _i}, \text{ } {\nabla _\theta }\mathcal{L}  \right)$ \cite{kingma2015}\;
		}
	}
\end{algorithm}
An offline supervised learning approach is adopted to train the PI-GRU network using pre-labeled samples containing measured TA temperatures and TEC currents under low-power safe conditions. As outlined in Algorithm \ref{alg.1}, the model's parameter set $\theta$ is iteratively optimized through backpropagation toward enhancing prediction accuracy. The joint loss function of the PI-GRU model integrates physics soft constraints, formulated as:
\begin{equation} \label{eq.10}
	\mathcal{L}\left( \theta  \right) = \left( {1 - \lambda } \right){\mathcal{L}_{{\text{data}}}}\left( \theta  \right) + \lambda {\mathcal{L}_{{\text{phy}}}}\left( \theta  \right)
\end{equation} 
where $\lambda$ is a hyperparameter balancing the data loss \(\mathcal{L}_{{\text{data}}}\left( \theta \right)\) and the physics loss \(\mathcal{L}_{{\text{phy}}}\left( \theta  \right)\).

The data loss \(\mathcal{L}_{{\text{data}}}\left( \theta \right)\) quantifies the discrepancy between the predicted temperature \(\hat T\) and experimentally measured data \(\overset{\lower0.5em\hbox{$\smash{\scriptscriptstyle\frown}$}}{T}\) via mean squared error (MSE), defined as:  
\begin{equation} \label{eq.11}
	{\mathcal{L}_{{\text{data}}}}\left( \theta  \right) = \frac{1}{{Bm}}\sum\limits_{j = 1}^B {\sum\limits_{l = 1}^m {{{\left( {\hat T_l^j - \overset{\lower0.5em\hbox{$\smash{\scriptscriptstyle\frown}$}}{T} _l^j} \right)}^2}} }
\end{equation} 
where $B$ denotes the batch size representing the number of sample groups processed in each training iteration. 

To enforce thermodynamic consistency in the predictions, a physics regularization term \(\mathcal{L}_{{\text{phy}}}\left( \theta  \right)\) is constructed as a soft constraint. This constraint systematically incorporates the thermal dynamics governing TEC-driven processes through three primary components:  
\begin{align} \label{eq.12}
	{Q_{\text{P}}} &= S{I_{{\text{ctl}}}}{T_{\text{c}}},  \\
	{Q_{\text{J}}} &= \frac{1}{2}I_{{\text{ctl}}}^2{R_{\text{int} }},\\
	{Q_{\text{C}}} &= K\left( {{T_{\text{h}}} - {T_{\text{c}}}} \right), 
\end{align} 
where Seebeck coefficient \(S = 8.76 \times {10^{ - 4}}{\text{ }}{{\text{V}} \mathord{\left/{\vphantom {{\text{V}} {\text{K}}}} \right. \kern-\nulldelimiterspace} {\text{K}}}\), internal resistance ${R_{\operatorname{int} }} = 6.09{\text{ }}\Omega $, heat transfer coefficient $K = 0.373{\text{ }}{{\text{W}} \mathord{\left/ {\vphantom {{\text{W}} {\text{K}}}} \right. \kern-\nulldelimiterspace} {\text{K}}}$, and the hot side temperature ${T_{\text{h}}} = \SI{27}{\celsius}$ are experimentally measured parameters. Following the relationships ${I_{{\text{ctl}}}} = I_l^j$ and ${T_{\text{c}}} = \hat T_l^j$ from the system characterized in Section \ref{sec2}, the lumped-parameter thermal model is given by:
\begin{equation} \label{eq.15}
	\frac{{d{T_{\text{c}}}}}{{dt}} = \frac{1}{{{m_{\text{e}}}c}}\left( Q_{\text{P}} - Q_{\text{J}} - Q_{\text{C}} \right)
\end{equation} 
where ${m_{\text{e}}} = 0.3{\text{ kg}}$ and $c = 800{\text{ }}{{\text{J}} \mathord{\left/ {\vphantom {{\text{J}} {\left( {{\text{kg}} \cdot {\text{K}}} \right)}}} \right. \kern-\nulldelimiterspace} {\left( {{\text{kg}} \cdot {\text{K}}} \right)}}$  are the equivalent mass and heat capacity. After discretization, the physics loss \(\mathcal{L}_{{\text{phy}}}\left( \theta  \right)\) for the PI-GRU model is formulated as:  
\begin{equation} \label{eq.16}
	\begin{aligned}
		{\mathcal{L}_{{\text{phy}}}}\left( \theta  \right) &= \frac{1}{{Bm}}\sum\limits_{j = 1}^B {\sum\limits_{l = 2}^m {\left( \frac{\hat T_l^j - \hat T_{l-1}^j}{\Delta t} - \frac{1}{m_\text{e}c}\left( {SI_l^j\hat T_l^j} \right. \right. } }  \\
		&\left. \left. { - \frac{1}{2}{{\left( {I_l^j} \right)}^2}{R_{\text{int} }} - K\left( {{T_{\text{h}}} - \hat T_l^j} \right)} \right) \right)^2.
	\end{aligned}
\end{equation} 

By embedding thermodynamic governing equations, the PI-GRU model implicitly encodes essential system properties, including thermal capacitance and heat conduction pathways. The approach secures compliance with thermodynamic principles during the prediction of controlled temperature, even in out-of-domain scenarios. Since the temporal evolution of temperature inherently encapsulates the whole information with external thermal management interactions and internal thermodynamic responses, the historical temperature data alone provide sufficient observability for learning the TA's nonlinear dynamics. The well-trained PI-GRU harnesses its deep nonlinear representation capacity to decode complex system characteristics, thereby performing physics-constrained predictions of temperature progression under prospective control strategies.
\subsection{PINN-enhanced MPC}
\label{sec3c}
\begin{algorithm}[!ht]
	
	\caption{Hierarchical Parallel MPC Architecture}
	\label{alg.2}
	\KwIn{Serial data steam $T$, temperature setpoint $T_\text{ref}$,}
	\KwOut{Optimal control action $I_1^*\left( k \right)$.}
	
	\LinesNumbered
	Initialization: ${\tilde x^p},{\text{ }}{\tilde x_q} \leftarrow 0$, ${\text{ }}{\tilde J^p},{\text{ }}{\tilde J_q} \leftarrow \infty $.
	
	Background thread:
	
	\While{$T$ available}
	{
		Update historical queue $\bar T$ with $T$\;
	}
	
	Main program:
	
	\For{$k = n,{\text{ }}n + 1,{\text{ }} \cdots $}
	{
		Generate particles in solution space ${\Omega _I}$ with random initial velocities $v_1^p\left( k \right)$ and positions $x_1^p\left( k \right)$\;
		\For{$q = 1,{\text{ }}2,{\text{ }} \cdots ,{\text{ }30}$}
		{
			\Parallel{$p = 1,{\text{ }}2,{\text{ }} \cdots ,{\text{ }100}$}
			{
				Estimate predictive TA temperature queue   using the trained PI-GRU model \eqref{eq.2} with $\bar T\left( k \right)$ and $x_q^p\left( k \right)$\;
				\If{$\hat T\left( k \right) \in {\Xi _T}$}
				{
					Compute cost $J_q^p\left( k \right)$ using \eqref{eq.17}\;
				}
				\Else
				{
					Punish $x_q^p\left( k \right)$ by letting $J_q^p\left( k \right) \leftarrow {\tilde J^p}\left( k \right) + 1$\;
				}
				Update the individual optimal position ${\tilde x^p}\left( k \right)$ by comparing $J_q^p\left( k \right)$ and ${\tilde J^p}\left( k \right)$\;
				Update the global optimal position ${\tilde x_q}\left( k \right)$ by comparing $J_q^p\left( k \right)$ and ${\tilde J_q}\left( k \right)$\;
				Update $v_{q + 1}^p\left( k \right)$ and $x_{q + 1}^p\left( k \right)$ of each particle using \eqref{eq.18}--\eqref{eq.19}\;
			}
		}
		Obtain the optimal solution ${I^*}\left( k \right) \leftarrow {\tilde x_{30}}\left( k \right)$ and implement $I_1^*\left( k \right)$\;
	}
\end{algorithm}
Due to the global validity of  \eqref{eq.2}, the trained PI-GRU network demonstrates comprehensive predictive capabilities across all sampling instants within the MPC control horizon. The optimal control sequence ${I^*}\left( k \right)$ defined in \eqref{eq.1} is obtained through constrained optimization that minimizes the specified cost function. Considering the essential control requirement of set-point regulation in the TA temperature control system, the cost function is formulated as: 
\begin{equation} \label{eq.17}
	J_q^p\left( k \right) = \sum\limits_{l = 1}^m {{{\tilde T}_l}{{\left( k \right)}^2}}  + 0.01\sum\limits_{l = 1}^m {\Delta {I_l}{{\left( k \right)}^2}}
\end{equation}  
where ${\tilde T_l}\left( k \right) = {\hat T_l}\left( k \right) - {T_{{\text{ref}}}}$ quantifies the tracking error and $\Delta {I_l}\left( k \right) = {I_l}\left( k \right) - {I_{l - 1}}\left( k \right)$ denotes the control input variation.

Nevertheless, the computational burden of receding horizon optimization poses a significant challenge for deploying deep neural network-based MPC strategy in time-critical closed-loop systems. Thus, we design a hierarchical parallel architecture comprising three layers, as outlined in Algorithm \ref{alg.2}.  

\subsubsection{Data Acquisition Layer} 
A dedicated dual-buffering mechanism manages real-time TA temperature data streams, mitigating memory contention between primary control routines and serial I/O operations. Moverover, an event-driven asynchronous I/O framework with non-blocking semantics continuously monitors serial port descriptors, ensuring lock-free concurrency between real-time data acquisition and MPC optimization processes.

\subsubsection{Prediction Layer}
The PI-GRU model leverages GPU-accelerated parallel computation to simultaneously generate multi-step temperature predictions $\bar T\left( k \right)$ through incorporation of the candidate TEC actuation profiles \(\bar I\left( k \right)\) and the extrapolative thermodynamics characteristics learned from historical temperature observations \(\hat T\left( k \right)\).

\subsubsection{Optimization Layer}
A constrained particle swarm optimization (PSO) paradigm is employed to circumvent local optima traps inherent in gradient-based solvers, significantly boosting convergence speed. Throughout the real-time predictive control scheme, each candidate control sequence $\bar I\left( k \right)$ constitutes a particle $x\left( k \right)$ evolving through the actuation-constrained feasible space ${\Omega _I} = \left\{ {\omega \in {\mathbb{R}^m}\mid {I_{\min }} \leqslant {\omega _i} \leqslant {I_{\max }},{\text{ }}\forall i \in \left\{ {1,{\text{ }}2,{\text{ }} \cdots ,{\text{ }}m} \right\}} \right\}$, where current saturation boundaries enforce physical actuator limitations. For particle \(p\) at iteration \(q\), its position \(x_q^p\left( k \right)\) and velocity \(v_q^p\left( k \right)\) evolution follows the projected dynamics:
\begin{align}
	\label{eq.18}
	v_{q + 1}^p\left( k \right) &= wv_q^p\left( k \right) + {c_1}{r_1}\left( {{{\tilde x}^p}\left( k \right) - x_q^p\left( k \right)} \right) \\
	&+ {c_2}{r_2}\left( {{{\tilde x}_q}\left( k \right) - x_q^p\left( k \right)} \right) \nonumber \\
	\label{eq.19}
	x_{q + 1}^p\left( k \right) &= {\mathcal{P}_{{\Omega _I}}}\left( {x_q^p\left( k \right) + v_{q + 1}^p\left( k \right)} \right)
\end{align}
where ${\tilde x^p}\left( k \right)$ and ${\tilde x_q}\left( k \right)$ respectively denote the individual particle's optimum and the swarm's global optimum positions, with inertia weight \(\omega=0.5\), acceleration constants \(c_1=c_2=1.5\), and random numbers \(r_1, r_2 \in \left[0,1\right]\). The projection operator $\mathcal{P}_{\Omega _I}$ enforces control input constraints ${\Omega _I}$, whereas thermal safety guarantees are achieved through checking constraints ${\Xi _T} = \left\{ { \xi \in {\mathbb{R}^m}\mid {T_{\min }} \leqslant {\xi _i} \leqslant {T_{\max }},{\text{ }}\forall i \in \left\{ {1,{\text{ }}2,{\text{ }} \cdots ,{\text{ }}m} \right\}} \right\}$ for all iterations.

The proposed PI-GRU MPC strategy offers dual advantages. 
\begin{enumerate}[(1)]
	\item In model accuracy, a hybrid modeling framework is established by integrating data-based methods with physics-based principles, constructing a predictive model enriched with physical prior knowledge. Compared to data-driven models, PI-GRU exhibits enhanced generalizability under out-of-domain testing scenarios, effectively suppressing control deviations caused by model mismatch.
	\item Regarding computational efficiency, PI-GRU MPC achieves full-process acceleration through a three-layer parallelized architecture. The data acquisition layer employs event-driven mechanisms for asynchronous task scheduling, eliminating resource waste inherent in traditional polling methods. The prediction layer accelerates multi-batch tensor computations of PI-GRU using GPU parallel architectures, significantly reducing inference time. The optimization layer replaces serial gradient solver with swarm intelligence algorithms for distributed particle evaluation, enabling parallel optimization. The synergy of these three layers ensures millisecond-level latency and substantially reduces hardware resource consumption.
\end{enumerate} 
\section{Experimental Investigation}\label{sec4}
This section presents experimental validation on the TA laser temperature control system depicted in Fig. \ref{fig1}. Predictive accuracy comparisons highlight the PI-GRU model's superior extrapolation capability to unseen operating conditions, whereas control performance evaluations reveal the robustness of the proposed hierarchical parallel MPC.

\subsection{Data Collection for the Training Set}\label{sec4.a}
The training set comprises approximately $20000 \text{ s}$ open-loop operational data of TEC current and TA temperature, sampled at $5\text{ Hz}$ across seven low-power safe operating points spanning $0.3{\text{ W}}$ to $0.6{\text{ W}}$ in $0.05{\text{ W}}$ increments. To capture comprehensive system dynamics, TEC current sequences are randomized within the full range of $ \pm 2 {\text{ A}}$, with randomized update frequencies ranging from $0.05{\text{ Hz}}$ to $5{\text{ Hz}}$. All experiments maintained safe TA chip temperatures between $\SI{5}{\celsius}$ and $\SI{40}{\celsius}$.  Architectural details and training configuration are provided in Table \ref{table1}.
\begin{table}[t]
	\caption{Model hyperparameters and Training Configurations}
	\label{table1}
	\centering
	\resizebox{0.65\columnwidth}{!}{
		\begin{tabular}{l l}
			\hline\hline 
			\multicolumn{1}{c}{Parameter} & \multicolumn{1}{c}{\pbox{20cm}{Configuration}}  \\
			\hline
			Historical sequence length $n$ & $100$\\
			Future prediction steps $m$& $5$\\
			Hyperparameter for balancing losses   
			$\lambda$& $0.001$\\
			Number of epochs $E$& $10$\\
			Batch size $B$& $32$\\
			Learning rate $\eta$& $0.01$\\
			Learning rate scheduler& StepLR\\
			Optimizer & Adam\\
			\hline\hline
		\end{tabular}
	}
\end{table}

\textit{Remark 1:} In deep learning applications, neural network architectures and training parameters require experimental validation rather than theoretical derivation. Whilst extended historical sequences theoretically offer richer system information, excessive sequence lengths beyond practical relevance thresholds only increase computational costs without improving accuracy. The saturation can be explained by weakening temporal dependencies in temperature control dynamics. To balance physical constraints and feature learning, we implement phased constraint introduction through epoch-progressive weighting. The training set size is empirically optimized through randomized TEC current sequences that maximize operational scenario diversity. Subsequent experimental analysis confirm the PI-GRU model's effectiveness in identifying TA system dynamics.
\begin{table}[h]
	\renewcommand{\arraystretch}{1.3}
	\caption{Quantitative Comparison of Predictive Performance between GRU and PI-GRU at Three Powers}
	\label{table2}	
	\centering
	\resizebox{\columnwidth}{!}{
		\begin{tabular}{c c ccc ccc}
			\hline\hline
			\multicolumn{2}{c}{Model} &
			\multicolumn{3}{c}{GRU} &
			\multicolumn{3}{c}{PI-GRU} \\
			\hline
			\multicolumn{2}{c}{Index} &
			MAE ($\SI{}{\celsius}$) & RMSE ($\SI{}{\celsius}$) & MAPE (\%) &
			MAE ($\SI{}{\celsius}$) & RMSE ($\SI{}{\celsius}$) & MAPE (\%) \\
			\hline
			
			\multirow{6}{*}{$0.5 \text{ W}$}
			& Step 1 & $0.0039$ & $0.0046$ & $0.0169$ & $0.0020$ & $0.0026$ & $0.0080$ \\
			& Step 2 & $0.0042$ & $0.0057$ & $0.0195$ & $0.0019$ & $0.0024$ & $0.0077$ \\
			& Step 3 & $0.0044$ & $0.0057$ & $0.0199$ & $0.0024$ & $0.0032$ & $0.0101$ \\
			& Step 4 & $0.0053$ & $0.0071$ & $0.0245$ & $0.0029$ & $0.0037$ & $0.0118$ \\
			& Step 5 & $0.0061$ & $0.0078$ & $0.0277$ & $0.0036$ & $0.0046$ & $0.0151$ \\
			& Overall & $0.0048$ & $0.0063$ & $0.0217$ & $0.0025$ & $0.0034$ & $0.0105$ \\
			
			\hline
			
			\multirow{6}{*}{1.0W}
			& Step 1 & $0.0064$ & $0.0076$ & $0.0279$ & $0.0025$ & $0.0035$ & $0.0098$ \\
			& Step 2 & $0.0070$ & $0.0095$ & $0.0323$ & $0.0028$ & $0.0039$ & $0.0111$ \\
			& Step 3 & $0.0072$ & $0.0095$ & $0.0327$ & $0.0030$ & $0.0041$ & $0.0118$ \\
			& Step 4 & $0.0088$ & $0.0118$ & $0.0407$ & $0.0037$ & $0.0051$ & $0.0145$ \\
			& Step 5 & $0.0100$ & $0.0130$ & $0.0459$ & $0.0044$ & $0.0059$ & $0.0173$ \\
			& Overall & $0.0079$ & $0.0105$ & $0.035$ & $0.0033$ & $0.0046$ & $0.0129$ \\
			
			\hline
			
			\multirow{6}{*}{1.5W}
			& Step 1 & $0.0067$ & $0.0079$ & $0.0291$ & $0.0018$ & $0.0024$ & $0.0073$ \\
			& Step 2 & $0.0076$ & $0.0098$ & $0.0345$ & $0.0015$ & $0.0020$ & $0.0058$ \\
			& Step 3 & $0.0077$ & $0.0099$ & $0.0348$ & $0.0029$ & $0.0035$ & $0.0122$ \\
			& Step 4 & $0.0093$ & $0.0121$ & $0.0427$ & $0.0032$ & $0.0040$ & $0.0135$ \\
			& Step 5 & $0.0106$ & $0.0133$ & $0.0478$ & $0.0044$ & $0.0053$ & $0.0186$ \\
			& Overall & $0.0084$ & $0.0108$ & $0.0378$ & $0.0028$ & $0.0036$ & $0.0115$ \\
			\hline\hline
		\end{tabular}
	}
\end{table}
\subsection{Prediction Performance and Ablation Studies}
\label{sec4b}
The multi-step prediction accuracy and extrapolation capability of the PI-GRU network are examined as critical prerequisites for MPC optimization. To investigate the contribution of thermodynamic knowledge integration, ablation studies are conducted by removing the physical constraints, resulting in a purely data-driven GRU model for comparison.

For comparative fairness, both PI-GRU and GRU models share identical training data using seven low-power datasets in Section \ref{sec4.a}. Table \ref{table2} illustrates performance comparisons of global error distribution analysis and stepwise error percentage tracking, quantitatively benchmarking both methods on test data using standard metrics: MAE (mean absolute error), RMSE (root mean square error), and MAPE (mean absolute percentage error). The test set contains newly acquired data from two out-of-domain scenarios.

\subsubsection{Case \MakeUppercase{\romannumeral 1}---Low-power Operation}
The first test derives from closed-loop control implementation at the pre-trained $0.5\text{ W}$ power setting. ``Closed-loop'' implies a new set of data collected under feedback mechanisms, within the trained power range. Comparative analysis with the baseline GRU model in Case \MakeUppercase{\romannumeral 1} demonstrates PI-GRU's enhanced prediction accuracy under known operating conditions.
\begin{figure}[!tb]
	\centering
	\subfloat[]
	{
		\begin{minipage}[h]{0.31\columnwidth}
			\centering
			\includegraphics[width=\textwidth]{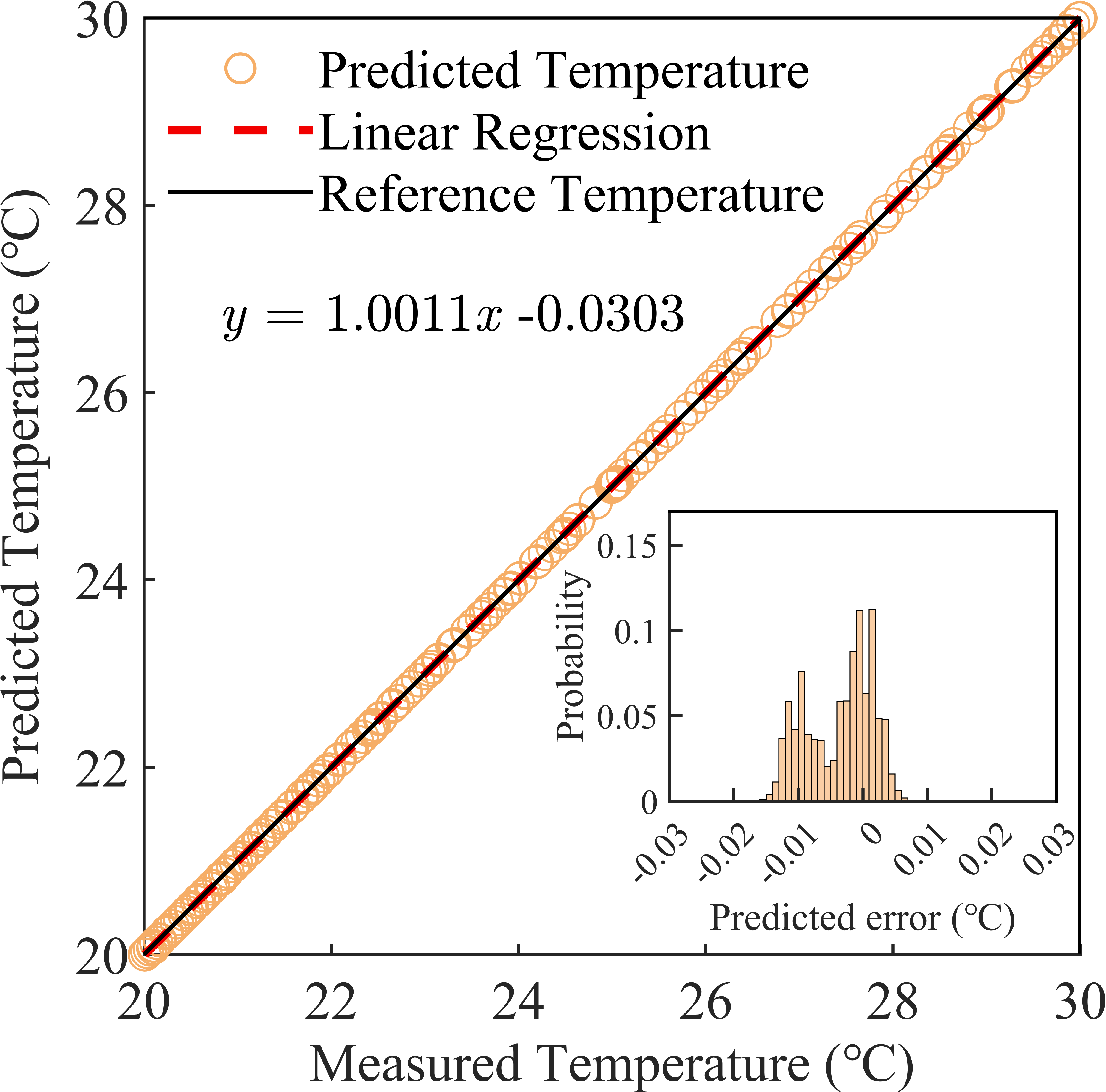}
			\vspace{-5mm}
			\label{fig4a}
		\end{minipage}
	}
	\subfloat[]
	{
		\begin{minipage}[h]{0.31\columnwidth}
			\centering
			\includegraphics[width=\textwidth]{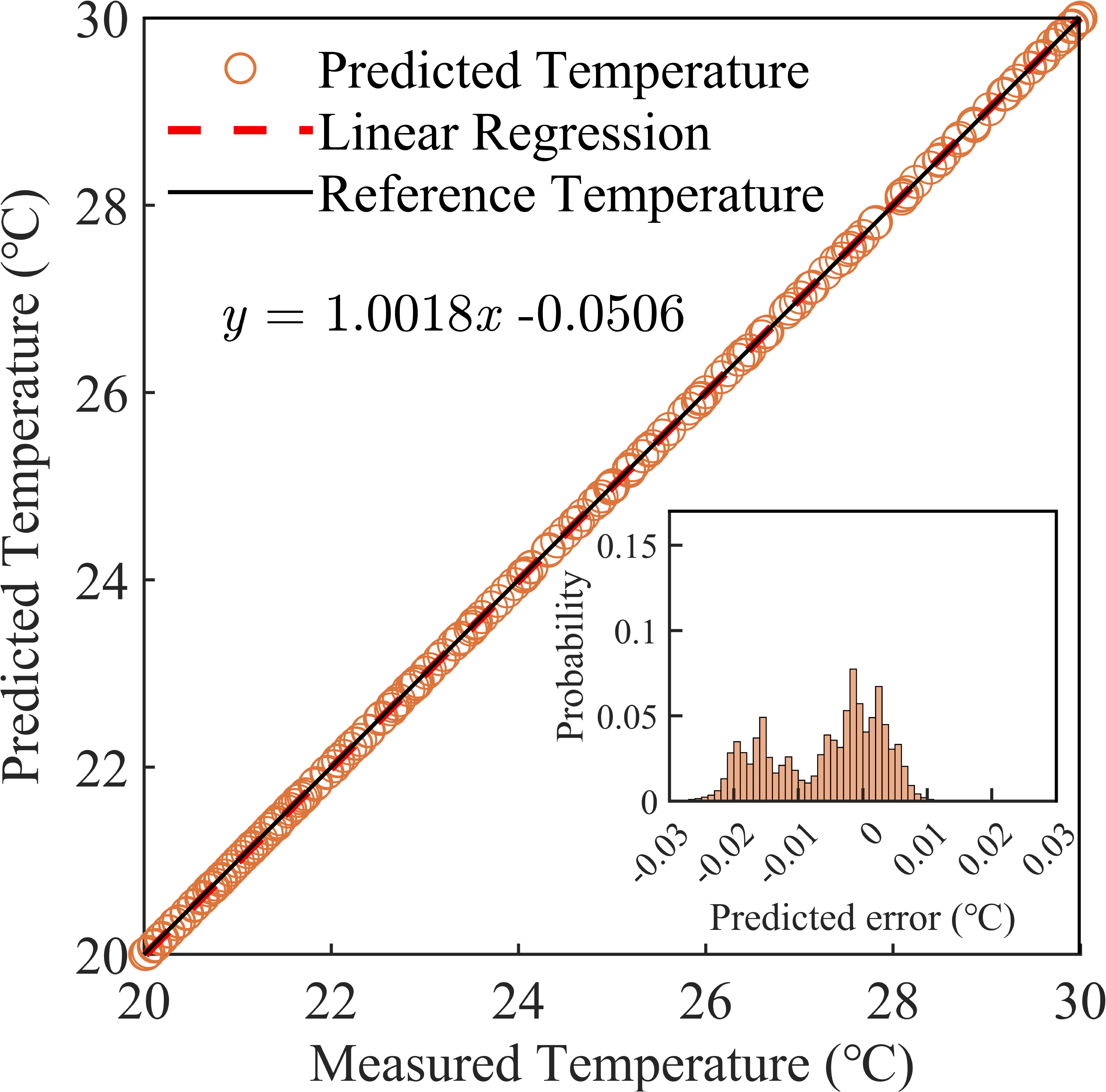}
			\vspace{-5mm}
			\label{fig4b}
		\end{minipage}	
	}	
	\subfloat[]
	{
		\begin{minipage}[h]{0.31\columnwidth}
			\centering
			\includegraphics[width=\textwidth]{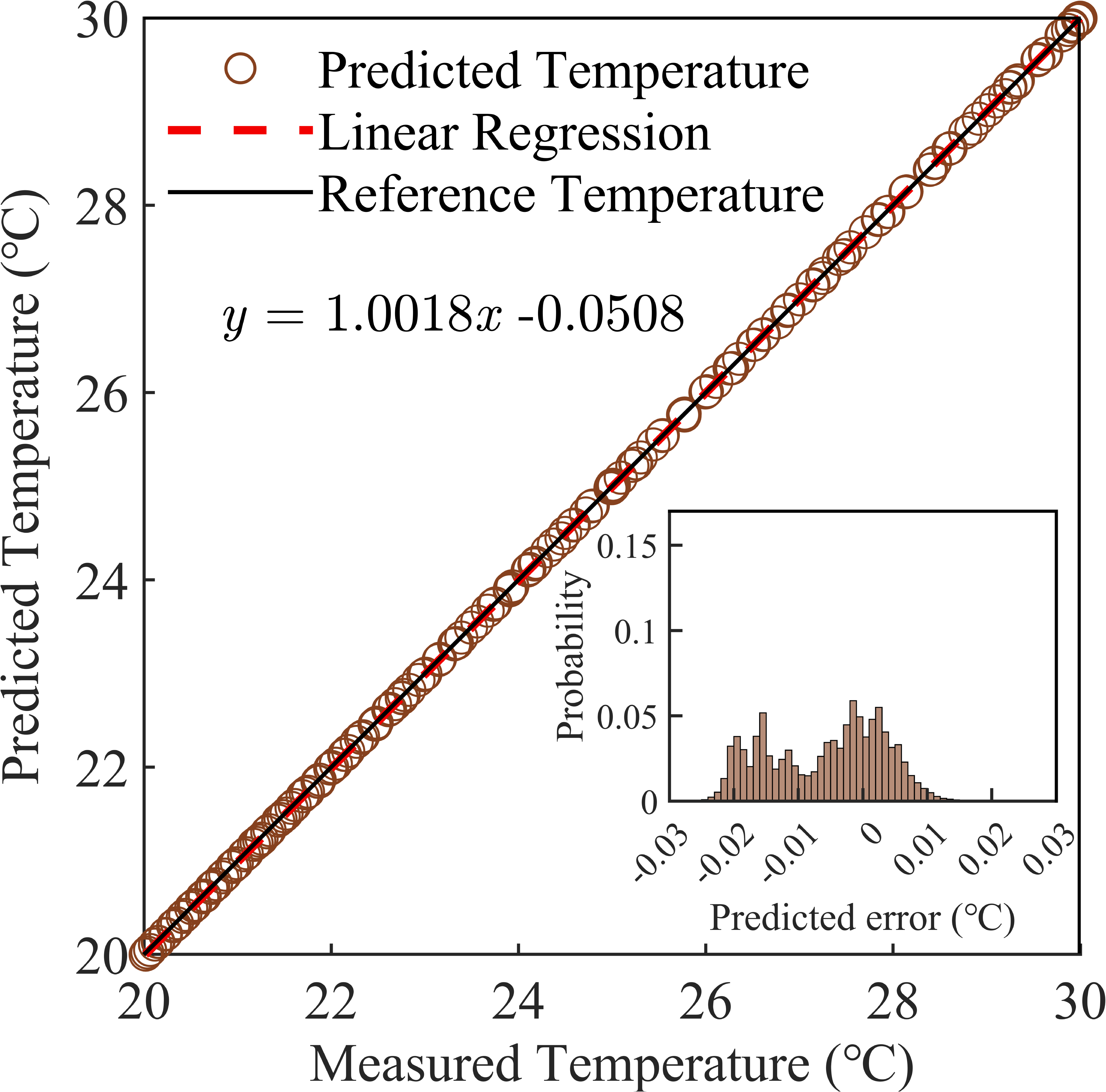}
			\vspace{-5mm}
			\label{fig4c}
		\end{minipage}
	}\\
	
	\subfloat[]
	{
		\begin{minipage}[h]{0.31\columnwidth}
			\centering
			\includegraphics[width=\textwidth]{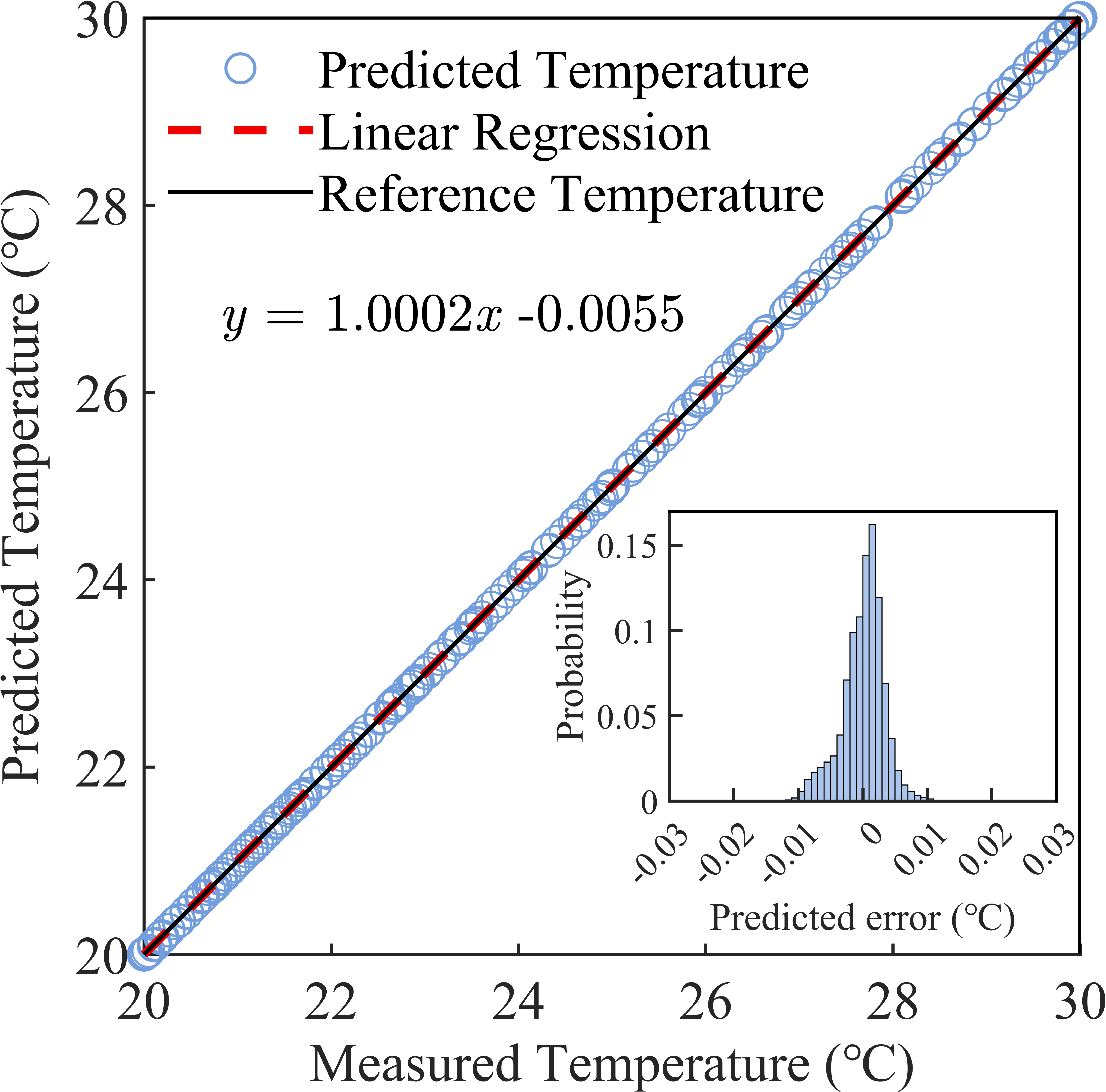}
			\vspace{-5mm}
			\label{fig4d}
		\end{minipage}
	}	
	\subfloat[]
	{
		\begin{minipage}[h]{0.31\columnwidth}
			\centering
			\includegraphics[width=\textwidth]{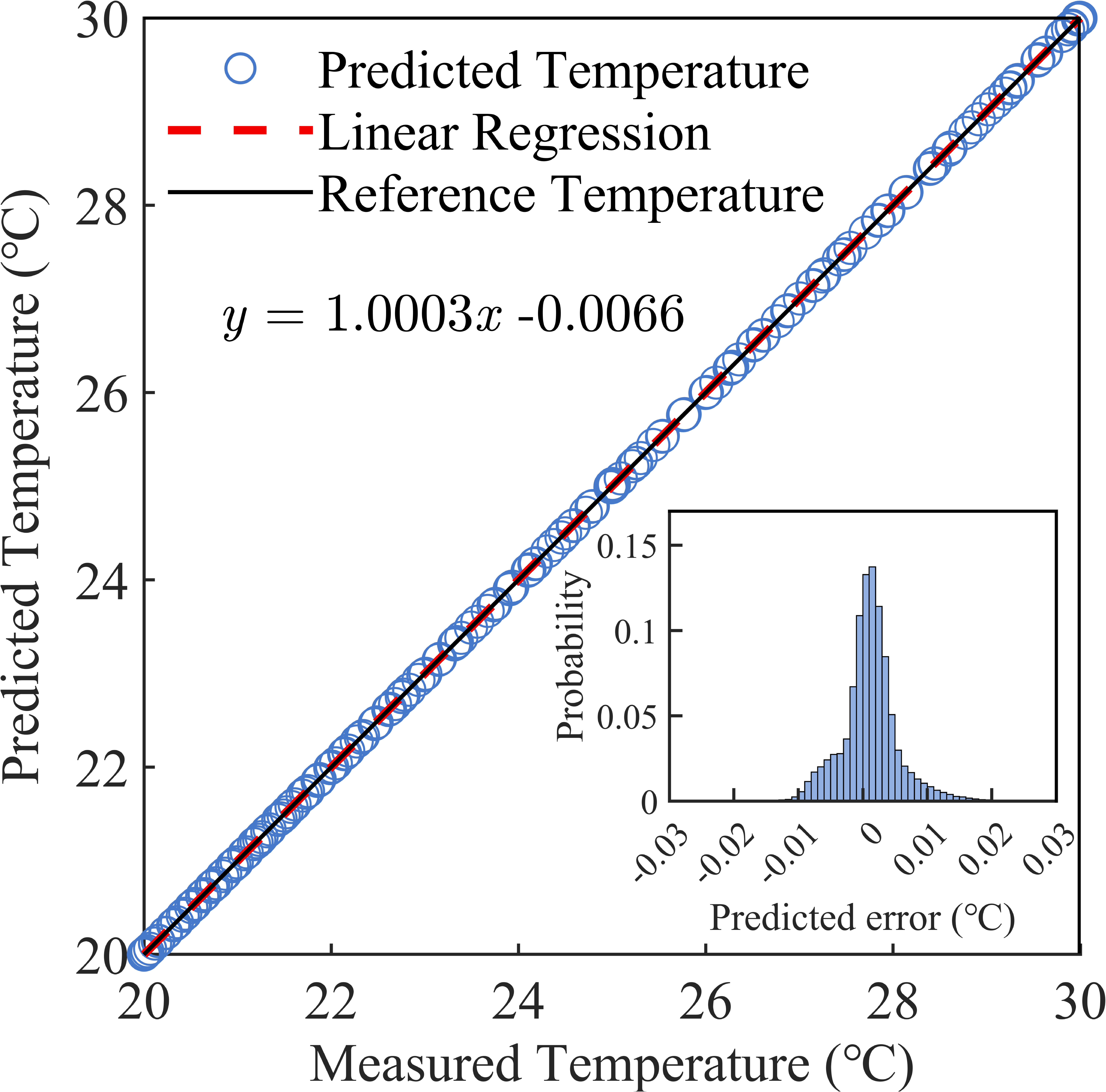}
			\vspace{-5mm}
			\label{fig4e}
		\end{minipage}
	}
	\subfloat[]
	{
		\begin{minipage}[h]{0.31\columnwidth}
			\centering
			\includegraphics[width=\textwidth]{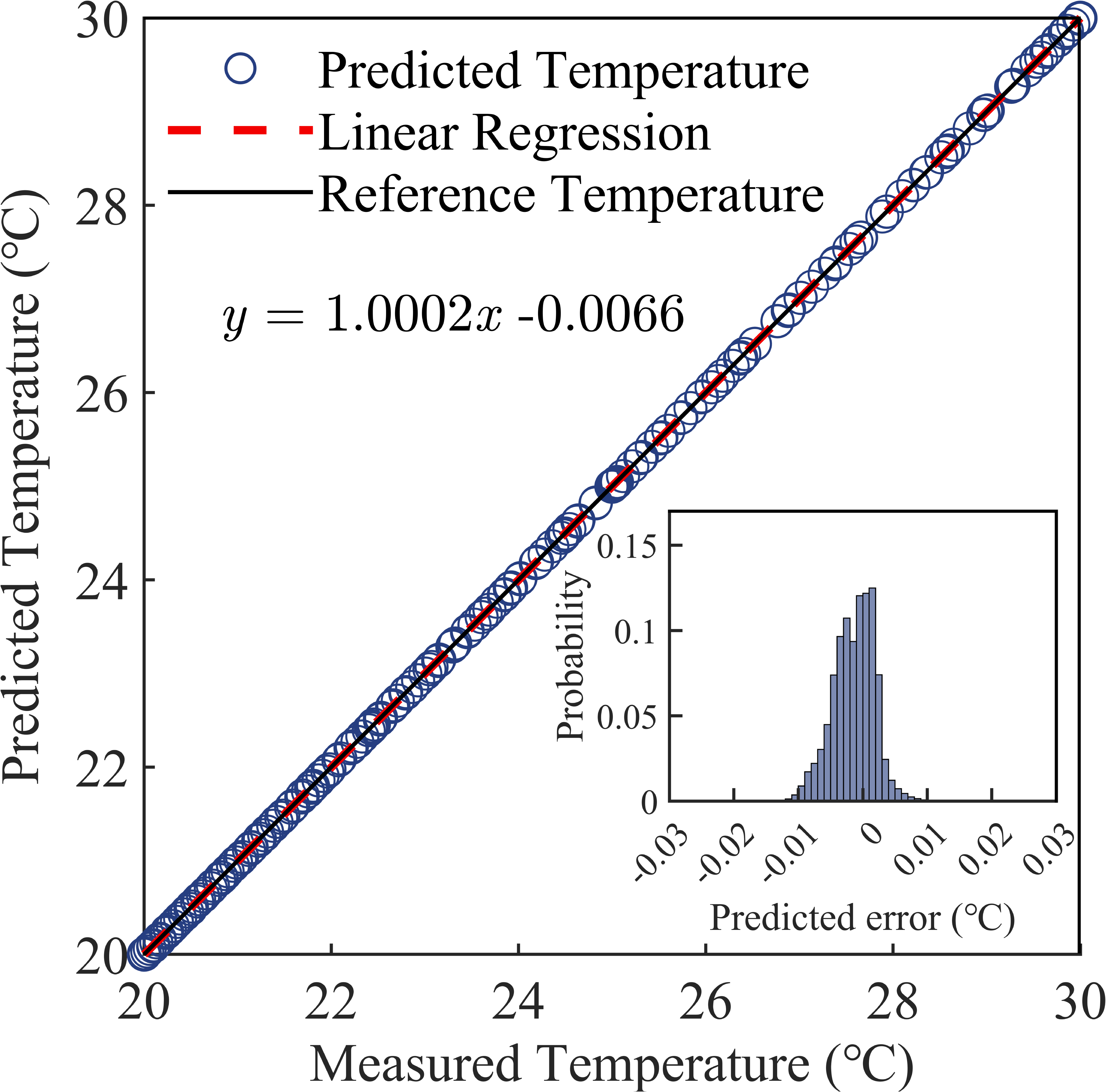}
			\vspace{-5mm}
			\label{fig4f}
		\end{minipage}
	}
	\caption{Prediction results of GRU and PI-GRU across three laser power operations, with subfigures showing probability histograms of prediction errors. (a) GRU at $0.5\text{ W}$. (b) GRU at $1\text{ W}$. (c) GRU at $1.5\text{ W}$. (d) PI-GRU at $0.5\text{ W}$. (e) PI-GRU at $1\text{ W}$. (f) PI-GRU at $1.5\text{ W}$.}
	\label{fig4}
\end{figure}

\begin{figure}[!tb]
	\centering
	\subfloat[]
	{
		\begin{minipage}[h]{0.85\columnwidth}
			\centering
			\includegraphics[width=\linewidth]{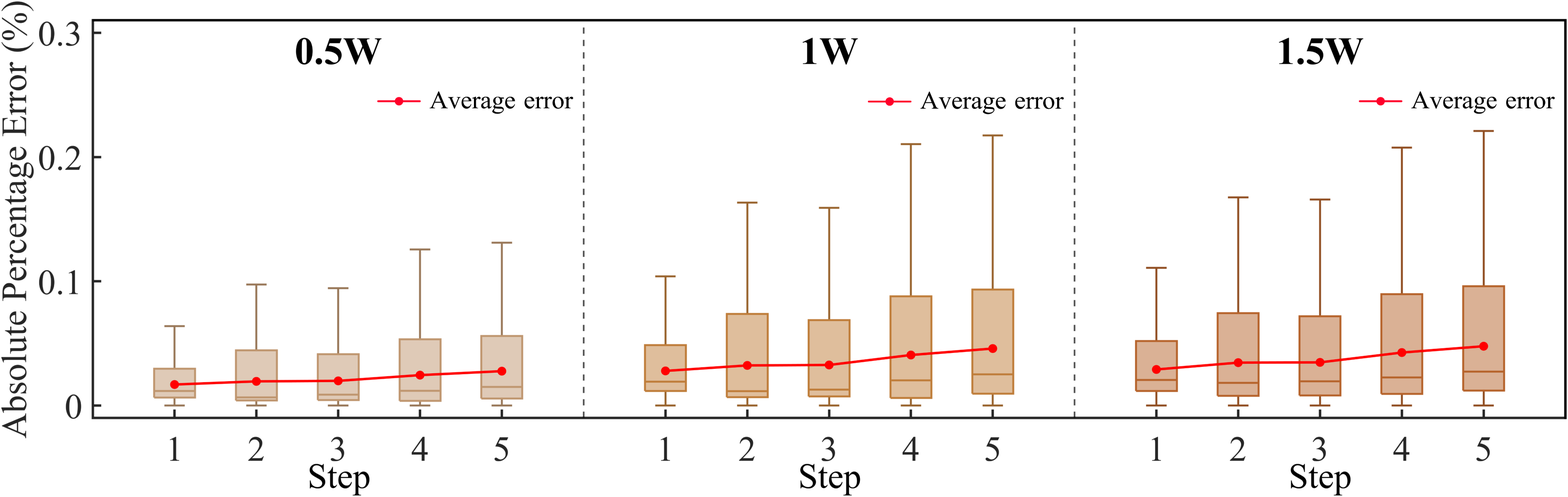}
			\vspace{-4mm}
			\label{fig5a}
		\end{minipage}
	}\\
	
	\subfloat[]
	{
		\begin{minipage}[h]{0.85\columnwidth}
			\centering
			\includegraphics[width=\linewidth]{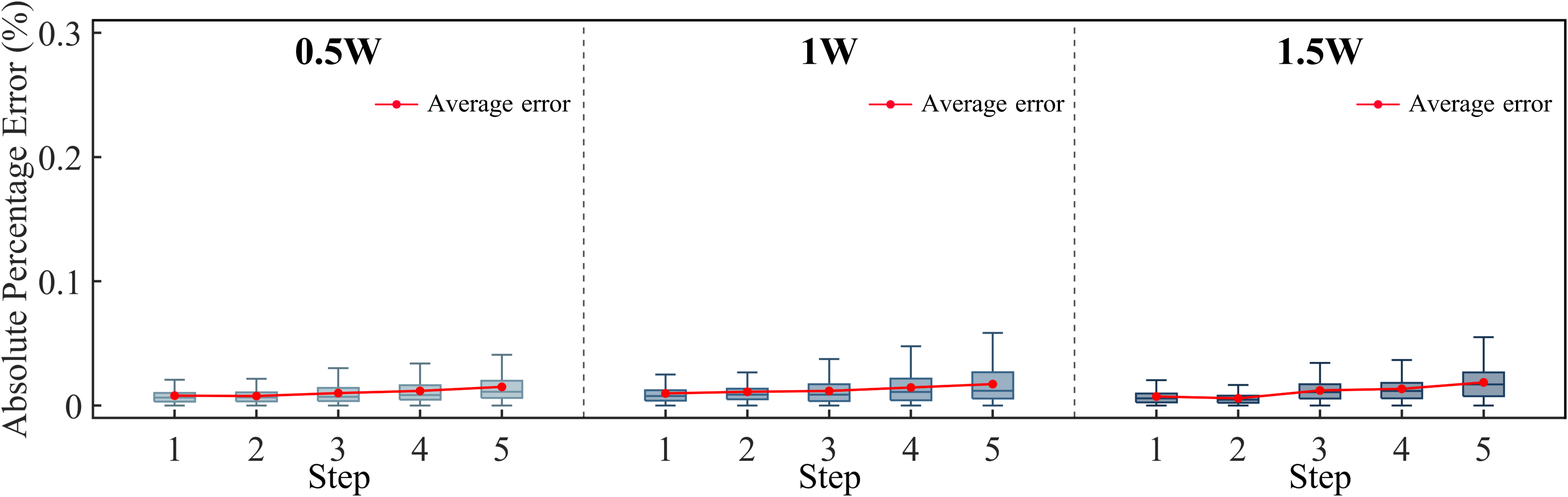}
			\vspace{-4mm}
			\label{fig5b}
		\end{minipage}	
	}
	\vspace{-1mm}
	\caption{Stepwise prediction errors of (a) GRU and (b) PI-GRU at three powers of $0.5\text{ W}$, $1\text{ W}$, and $1.5\text{ W}$.}
	\label{fig5}
\end{figure}

Fig. \ref{fig4}\subref{fig4a} and \subref{fig4c} present error distribution profiles of PI-GRU and GRU models in Case \MakeUppercase{\romannumeral 1}. PI-GRU achieves superior error confinement, with over $95\%$ of predictions falling within $\pm \SI{0.01}{\celsius}$, versus GRU's dispersion from $\SI{-0.03}{\celsius}$ to $\SI{0.01}{\celsius}$. Meanwhile, regression analysis reveals PI-GRU's near-ideal unity slope, confirming enhanced prediction fidelity. The stepwise errors in Fig. \ref{fig5} further show that PI-GRU maintains a consistent deviation of around $0.01\%$ across prediction horizons, whereas GRU exhibits a higher rate of $0.02\%$ with wider dispersion. Particularly, with notable reductions ($48.7\%$ MAE, $43.5\%$ RMSE, and $52.7\%$ MAPE) at the initial step, the PI-GRU's predictions provide critical accuracy advantages for subsequent control inputs evaluation.

These advancements originate from the physics regularization loss rooted in thermodynamic principles, systematically enhancing learned correlations between control inputs and temperature variations. By coupling the conduction model with the thermal regulation, the PI-GRU network captures constrained transient dynamics and suppresses steady-state errors. Technically, the physics-informed method outperforms data-driven counterparts solely relying on statistical correlations.
\subsubsection{Case \MakeUppercase{\romannumeral 2}---High-power Operation}
Whilst Case \MakeUppercase{\romannumeral 1} validates the PI-GRU's predictive effectiveness, its generalizability remains unverified. Accordingly, Case \MakeUppercase{\romannumeral 2} employs two unseen high-power conditions of $1\text{ W}$ and $1.5\text{ W}$, generating two new closed-loop test datasets through identical experimental setups. Furthermore, the evaluation methodology replicates Case \MakeUppercase{\romannumeral 1}'s configuration, thereby enabling a direct comparison of prediction accuracy across different power operations.

Compared to performance at $0.5\text{ W}$, the data-driven GRU exhibits unstable predictions beyond training domains, with global errors exceeding $\SI{0.04}{\celsius}$, as shown in Fig. \ref{fig4}\subref{fig4b} and \subref{fig4c}. Prediction accuracy deteriorates due to error propagation, particularly at steps 4 and 5, where outliers above $0.2\%$ are observed at $1 \text{ W}$ and $1.5 \text{ W}$ conditions, as illustrated in Fig. \ref{fig5}\subref{fig5a}. By contrast, the PI-GRU model sustains robust accuracy under out-of-domain high-power regimes, constraining the global temperature errors within $\SI{0.02}{\celsius}$ in Fig. \ref{fig4}\subref{fig4e} and \subref{fig4f} maintaining the stepwise percentage errors consistently below $0.05\%$ across all prediction horizons in Fig. \ref{fig5}\subref{fig5b}. Moreover, metrics in Table \ref{table2} quantify PI-GRU's enhanced extrapolation capability, with $58.2\%$ and $66.7\%$ MAE reduction and $56.2\%$ and $66.7\%$ RMSE reduction at $1\text{ W}$ and $1.5\text{ W}$ respectively, compared to GRU. Notably, PI-GRU achieves critical MPC-oriented improvements at the initial prediction step, reducing MAE by $60.9\%$ and $73.1\%$ and RMSE by $53.9\%$ and $69.6\%$ versus GRU under extrapolation scenarios.  

Experimental validations across Cases \MakeUppercase{\romannumeral 1} and \MakeUppercase{\romannumeral 2} highlight the PI-GRU model's  predictive reliability across three power conditions, thereby verifying generalizability beyond training specifications. Fundamentally, the superior inference mechanism stems from the embedded thermodynamic soft constraints governing solution spaces to physically admissible trajectories. This consistent accuracy establishes essential foundations for MPC integration, where temporal error thresholds dictate system stability. Conversely, conventional data-driven architectures exhibit inherent fragility to error accumulation, as evidenced by their divergence under high-power scenarios. Subsequent investigations evaluate the control performance enhancement via PINN integration.

\textit{Remark 2:} Note that safety concerns prohibit open-loop operation at high power for  lasers, prompting experimental evaluation of prediction performance using closed-loop data with feedback control. Whilst not as randomized as the training data, the test sets fully encompass the TA system's thermal dynamics, including transient responses and steady-state conditions. Significantly, test protocols incorporate setpoints spanning the $\SI{20}{\celsius}$--$\SI{30}{\celsius}$ operational range essential for TA lasers in SERF atomic magnetometer arrays. Therefore, the PI-GRU's demonstrated superiority in prediction accuracy and generalizability establishes practical significance for advancing laser quality in quantum sensing systems.
\begin{figure}[tbp]
	\centering
	\subfloat[]
	{
		\begin{minipage}[h]{0.8\columnwidth}
			\centering
			\includegraphics[width=\linewidth]{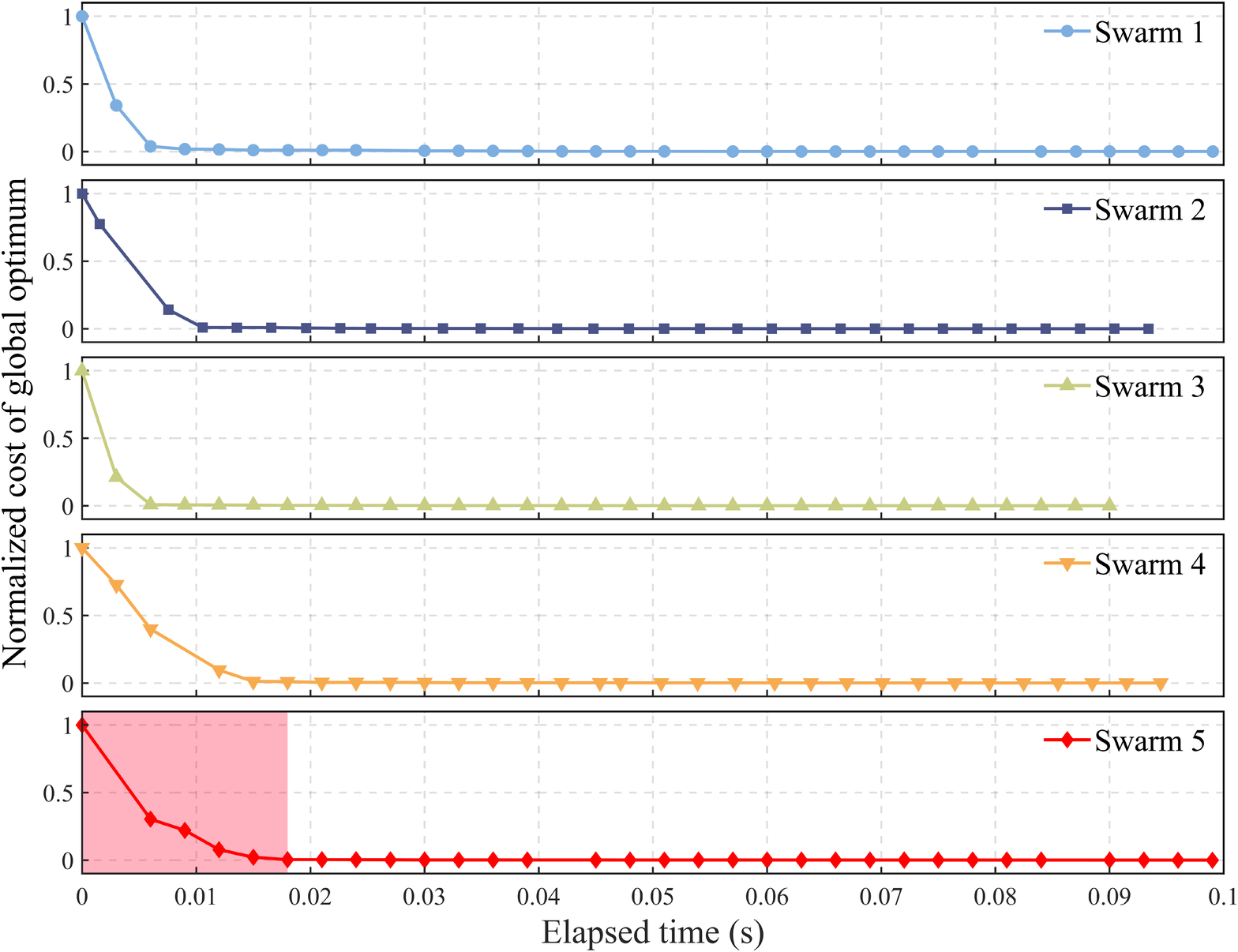}
			\vspace{-5mm}
			\label{fig6a}
		\end{minipage}
	}\\
	\subfloat[]
	{
		\begin{minipage}[h]{0.85\columnwidth}
			\centering
			\includegraphics[width=\linewidth]{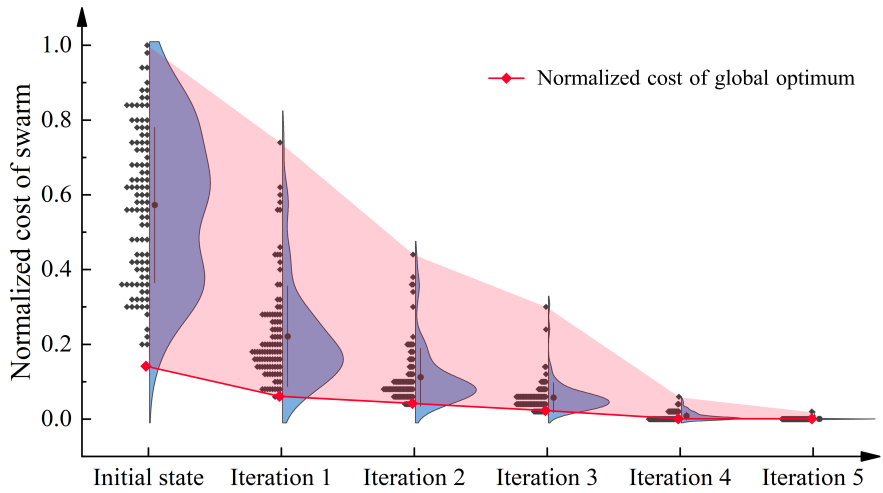}
			\vspace{-5mm}
			\label{fig6b}
		\end{minipage}	
	}
	\caption{Verification of the computational efficiency: (a) Evolutionary trajectory of the global optimum showing the convergence behavior of five swarms initialized with different states toward the stable region. (b) Detailed visualization of the particle aggregation process for the fifth group during convergence.}
	\vspace{-8mm}
	\label{fig6}
\end{figure}
\subsection{Control Performance}
\label{sec4c}
Based on the multi-step TA temperature trajectory predicted by the PI-GRU model, we first verify the computational efficiency of MPC implementation following Algorithm \ref{alg.2} in practical deployment. Through random sampling five particle swarms with distinct initial distributions, their optimization processes reveal that PSO iterations can be stably completed within $0.1{\text{ s}}$, i.e. half of the prescribed control cycle, thereby satisfying real-time constraints for closed-loop control, as shown in Fig. \ref{fig6}\subref{fig6a}. Furthermore, the global best cost decays to below $0.01\%$ of its initial magnitudes within the first $5$ iterations, a phase accounting for merely one-third of the total computation time. The rapid convergence without sacrificing optimality is further visualized through the particle aggregation patterns shown in Fig. \ref{fig6}\subref{fig6b}, where initially dispersed particles progressively migrate towards refined solution regions.

Next, the control performance is evaluated through $1000 \text{ s}$ steady-state data of TA temperature. To comprehensively assess the robustness of control strategies, nine test scenarios are conducted by combining three target temperatures ($\SI{20}{\celsius}$, $\SI{25}{\celsius}$, $\SI{30}{\celsius}$) with three power operations ($0.5\text{ W}$, $1\text{ W}$, $1.5\text{ W}$). In accordance with receding horizon optimization, the TEC control adjustment is synchronized with the data acquisition at a frequency of $5\text{ Hz}$.

\subsubsection{Performance Evaluation Metrics}
\begin{enumerate}[(1)]
	\item Range $R$ quantifies the maximum deviation amplitude:
	\begin{equation} \label{eq.20}
		R = \max _{k = 1}^N\left\{ {T\left( k \right)} \right\} - \min _{k = 1}^N\left\{ {T\left( k \right)} \right\}
	\end{equation}
	where \( N \) denotes the total count of data points.
	\item Standard deviation $\sigma$ quantifies the distribution of overall experimental data:
	\begin{equation} \label{eq.21}
		\sigma  = \sqrt {\frac{1}{N}\sum\limits_{k = 1}^N {{{\left( {T\left( k \right) - {T_{{\text{avg}}}}} \right)}^2}} }
	\end{equation}
	where ${T_{{\text{avg}}}}$ is the mean of all data.
	\item Allan deviation $\sigma \left( \tau  \right)$ provides a time-domain characterization of stochastic fluctuations by quantifying the temperature stability across varying averaging intervals: 
	\begin{equation} \label{eq.22}
		\sigma \left( \tau  \right) = \sqrt {\frac{1}{{2\left( {{N_\tau } - 1} \right)}}\sum\limits_{{k_\tau } = 1}^{{N_\tau } - 1} {{{\left( {{T_{{\text{avg}}}}\left( {{k_\tau } + 1} \right) - {T_{{\text{avg}}}}\left( {{k_\tau }} \right)} \right)}^2}} }
	\end{equation}	
	where ${N_\tau}$ represents the number of time blocks when partitioning the time series data into segments of length \( \tau \) and ${T_{{\text{avg}}}}\left( {{k_\tau }} \right)$ denotes the ${k_\tau }$-th block's average value. 
\end{enumerate}

\begin{table}[t]
	\renewcommand{\arraystretch}{1.3}
	\caption{Quantitative Comparison of Control Stability between three strategies under Nine Test Scenarios}
	\centering
	\label{table3}
	\resizebox{\columnwidth}{!}{
		\begin{tabular}{lcccccccccc}
			\hline\hline \\[-4mm]
			\multirow{2}{*}{Method} & 
			\multirow{2}{*}{Index} & 
			\multicolumn{3}{c}{$0.5 \text{ W}$} & 
			\multicolumn{3}{c}{$1 \text{ W}$} & 
			\multicolumn{3}{c}{$1.5 \text{ W}$} \\
			& & $\SI{20}{\celsius}$ & $\SI{25}{\celsius}$ & $\SI{30}{\celsius}$ & $\SI{20}{\celsius}$ & $\SI{25}{\celsius}$ & $\SI{30}{\celsius}$ & $\SI{20}{\celsius}$ & $\SI{25}{\celsius}$ & $\SI{30}{\celsius}$ \\
			\hline
			
			\multirow{5}{*}{LMPC}
			& $R$ ($\SI{}{\celsius}$) & $0.053$ & $0.051$ & $0.046$ & $0.055$ & $0.052$ & $0.069$ & $0.067$ & $0.062$ & $0.062$ \\
			& $\sigma$ ($\SI{}{\celsius}$) & $0.0095$ & $0.0093$ & $0.0090$ & $0.0105$ & $0.0089$ & $0.0129$ & $0.0125$ & $0.0107$ & $0.0127$ \\
			& $\sigma\left(1\text{ s}\right)$ ($\times10^{-3}\SI{}{\celsius}$) & $3.0018$ & $4.6363$ & $5.0989$ & $2.9057$ & $3.5598$ & $8.1348$ & $3.3561$ & $3.9840$ & $7.5867$ \\
			& $\sigma\left(10\text{ s}\right)$ ($\times10^{-3}\SI{}{\celsius}$) & $3.9486$ & $3.6908$ & $2.2397$ & $2.4849$ & $1.9600$ & $2.2487$ & $3.7614$ & $2.4805$ & $2.2091$ \\
			& $\sigma\left(100\text{ s}\right)$ ($\times10^{-3}\SI{}{\celsius}$) & $2.1186$ & $1.9358$ & $0.9545$ & $2.5427$ & $1.3531$ & $2.9228$ & $2.9657$ & $2.7663$ & $1.6364$ \\
			
			\hline
			
			\multirow{5}{*}{GRU MPC} 
			& $R$ ($\SI{}{\celsius}$) & $0.011$ & $0.010$ & $0.010$ & $0.017$ & $0.013$ & $0.014$ & $0.019$ & $0.017$ & $0.018$ \\
			& $\sigma$ ($\SI{}{\celsius}$) & $0.0022$ & $0.0016$ & $0.0020$ & $0.0039$ & $0.0028$ & $0.0026$ & $0.0040$ & $0.0033$ & $0.0032$ \\
			& $\sigma\left(1\text{ s}\right)$ ($\times10^{-3}\SI{}{\celsius}$) & $0.6305$ & $0.6841$ & $1.0368$ & $1.1188$ & $1.2033$ & $1.4588$ & $0.7762$ & $0.9540$ & $0.9741$ \\
			& $\sigma\left(10\text{ s}\right)$ ($\times10^{-3}\SI{}{\celsius}$) & $0.4631$ & $0.3892$ & $0.3628$ & $0.4304$ & $0.3973$ & $0.4431$ & $0.2768$ & $0.5062$ & $0.4434$ \\
			& $\sigma\left(100\text{ s}\right)$ ($\times10^{-3}\SI{}{\celsius}$) & $0.0707$ & $0.0981$ & $0.0746$ & $0.1435$ & $0.1254$ & $0.0679$ & $0.0516$ & $0.1223$ & $0.1066$ \\
			\hline
			
			\multirow{5}{*}{PI-GRU MPC} 
			& $R$ ($\SI{}{\celsius}$) & $0.007$ & $0.005$ & $0.007$ & $0.005$ & $0.006$ & $0.008$ & $0.008$ & $0.005$ & $0.007$ \\
			& $\sigma$ ($\SI{}{\celsius}$) & $0.0011$ & $0.0009$ & $0.0009$ & $0.0012$ & $0.0012$ & $0.0011$ & $0.0012$ & $0.0009$ & $0.0011$ \\
			& $\sigma\left(1\text{ s}\right)$ ($\times10^{-3}\SI{}{\celsius}$) & $0.3447$ & $0.3710$ & $0.3824$ & $0.3546$ & $0.3737$ & $0.4148$ & $0.3585$ & $0.3390$ & $0.4164$ \\
			& $\sigma\left(10\text{ s}\right)$ ($\times10^{-3}\SI{}{\celsius}$) & $0.2547$ & $0.2114$ & $0.1995$ & $0.2697$ & $0.2527$ & $0.2950$ & $0.2910$ & $0.2612$ & $0.2509$ \\
			& $\sigma\left(100\text{ s}\right)$ ($\times10^{-3}\SI{}{\celsius}$) & $0.0348$ & $0.0492$ & $0.0327$ & $0.0667$ & $0.0460$ & $0.0738$ & $0.0353$ & $0.0373$ & $0.0368$ \\
			\hline\hline
		\end{tabular}
	}
\end{table}

\subsubsection{Experimental Results Comparison}
To illustrate the control improvement through predictive model enhancement, three MPC strategies utilizing different models are implemented:
\begin{enumerate}[(1)]
	\item Linear MPC: The discrete predictive model is given by $T\left( k \right) = 0.028I\left( k \right) + 0.988T\left( {k - 1} \right)$, derived via linear regression on training set data.
	\item GRU MPC: The data-driven benchmark.
	\item PI-GRU MPC: The proposed PINN-enhanced strategy.  
\end{enumerate}
\begin{figure}[t]
	\centering
	\includegraphics[width=0.6\textwidth]{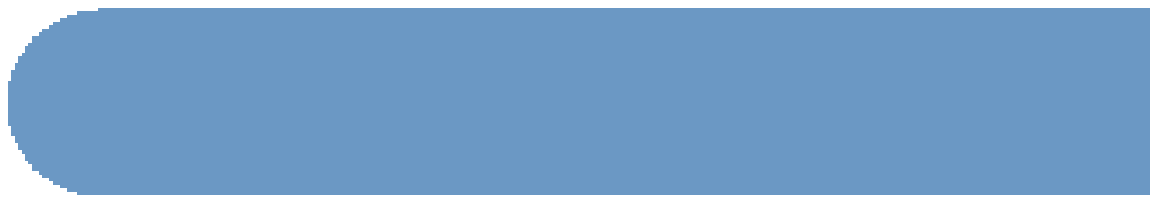}
	
	\newcommand{\twoplotrow}[2]{
		\subfloat[]
		{
			\begin{minipage}[t]{0.33\columnwidth}
				\centering
				\begin{minipage}[t]{0.56\columnwidth}
					\includegraphics[height=0.1\textheight]{#1}
				\end{minipage}
				\begin{minipage}[t]{0.40\columnwidth}
					\includegraphics[height=0.1\textheight]{#2}
				\end{minipage}
			\end{minipage}%
		}
	}
	
	\twoplotrow{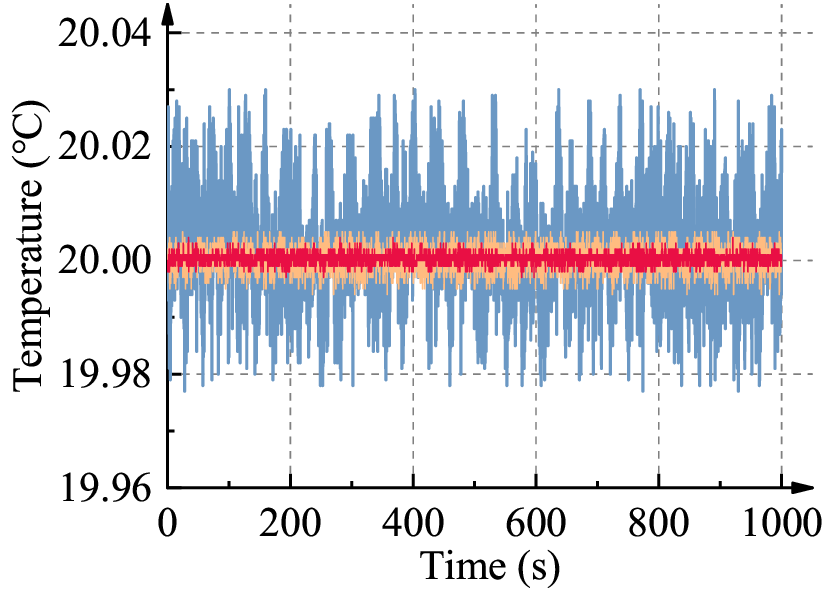}{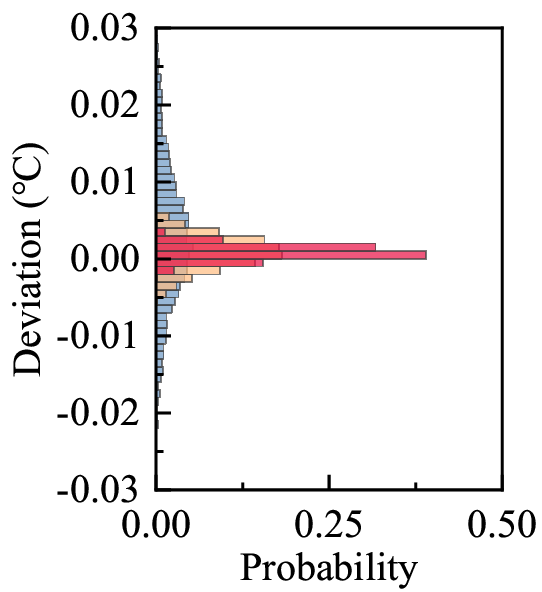}
	\twoplotrow{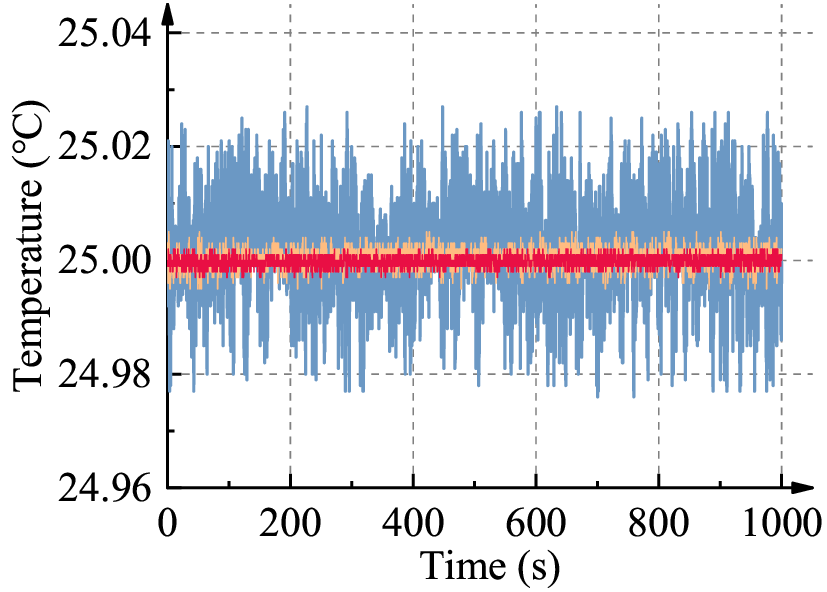}{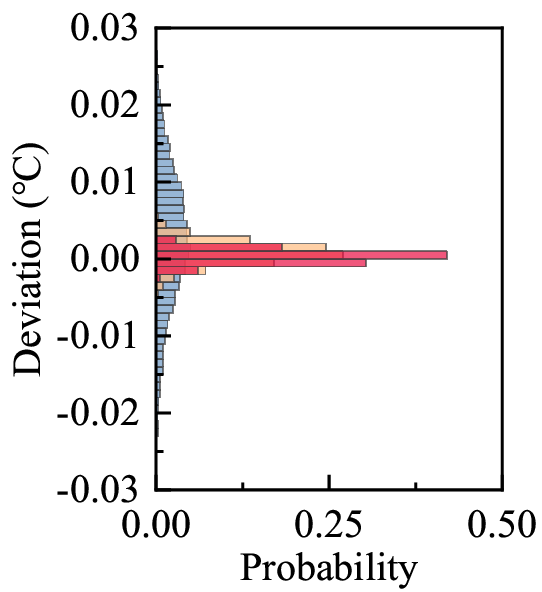}
	\twoplotrow{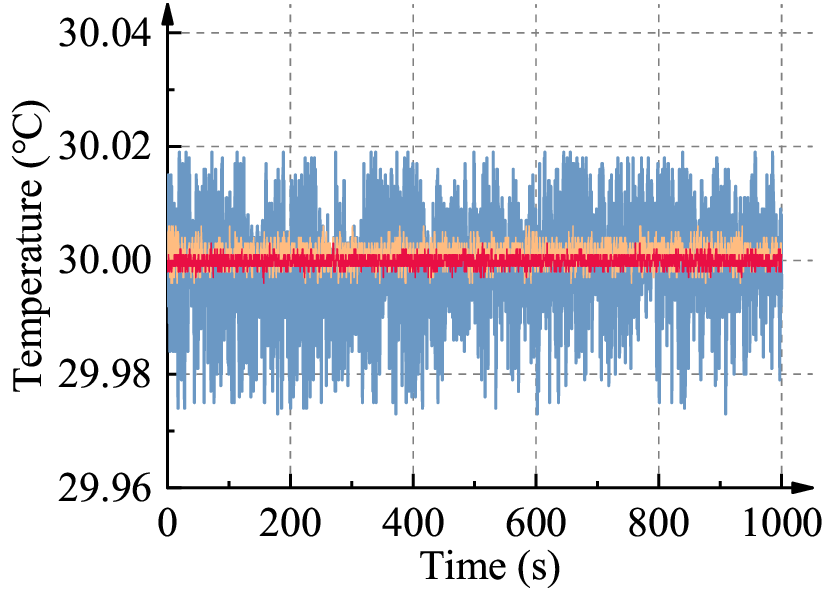}{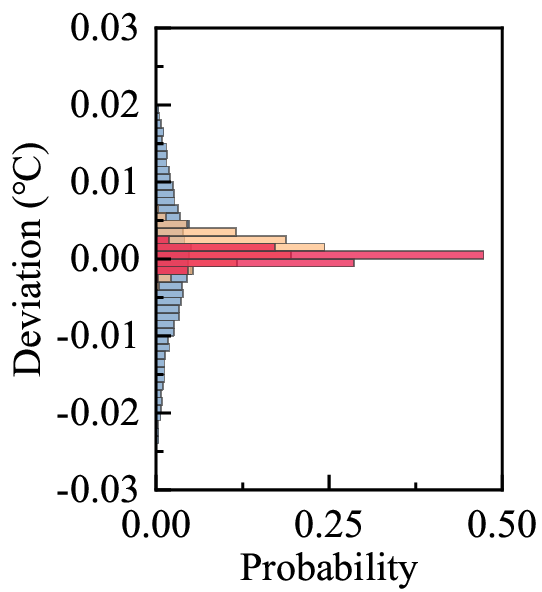}
	\\
	\vspace{-3mm}	
	
	\twoplotrow{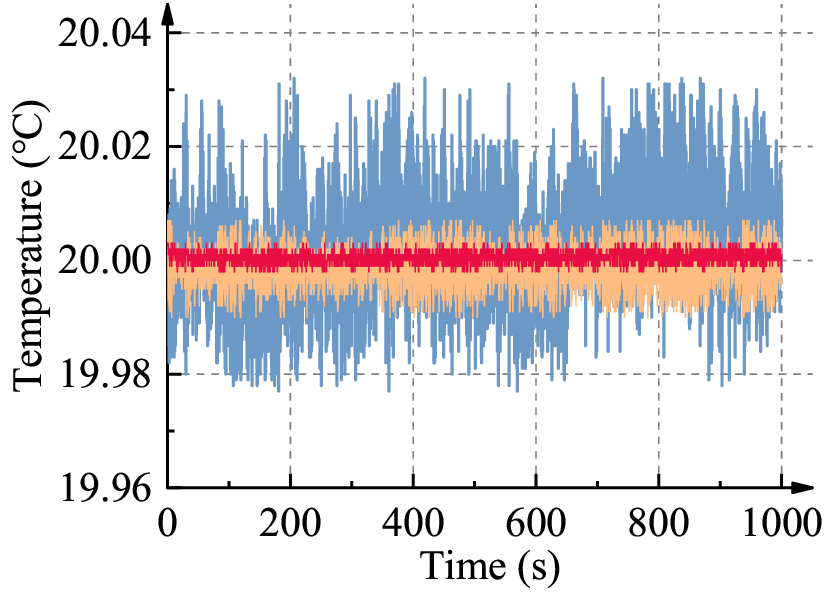}{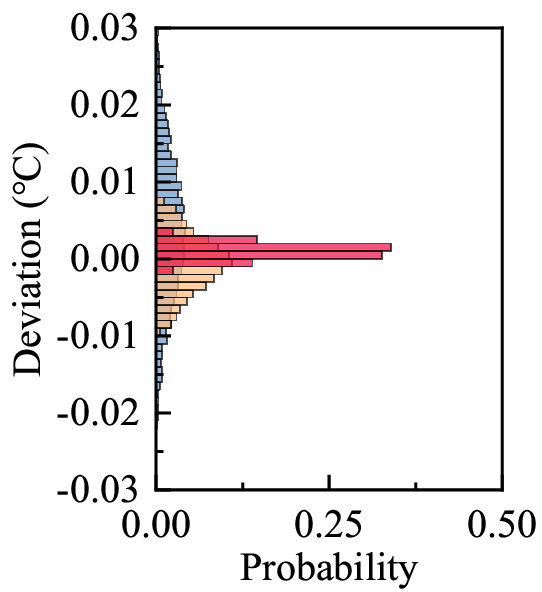}
	\twoplotrow{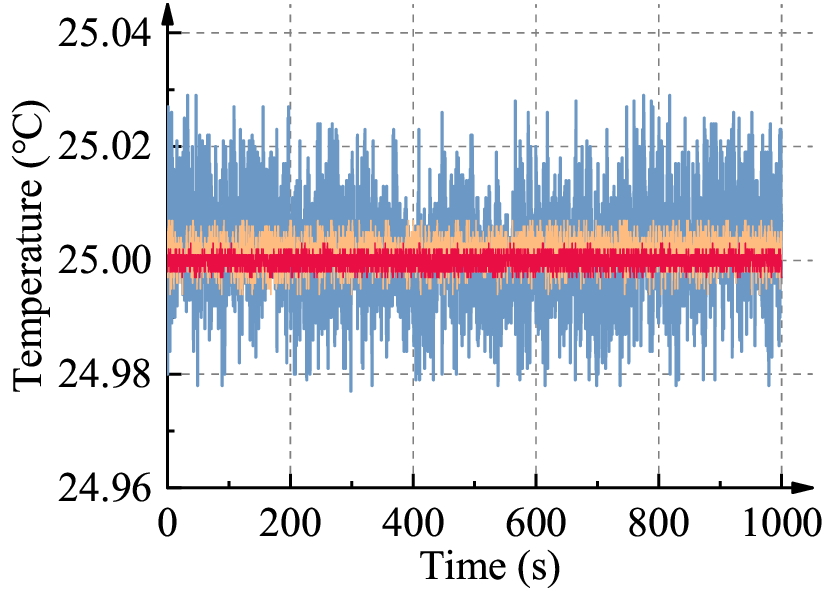}{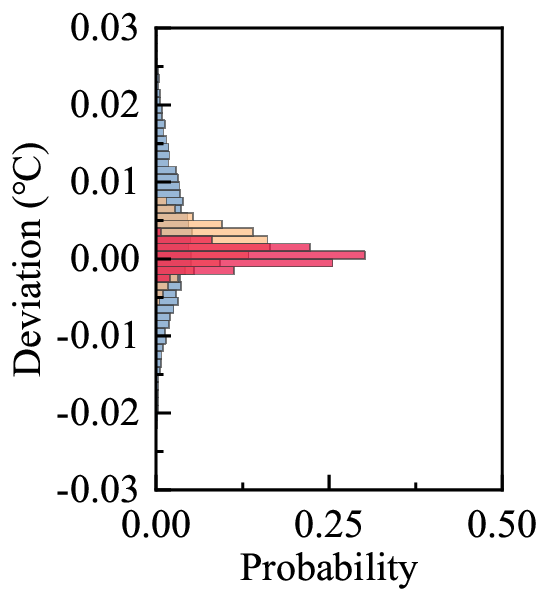}
	\twoplotrow{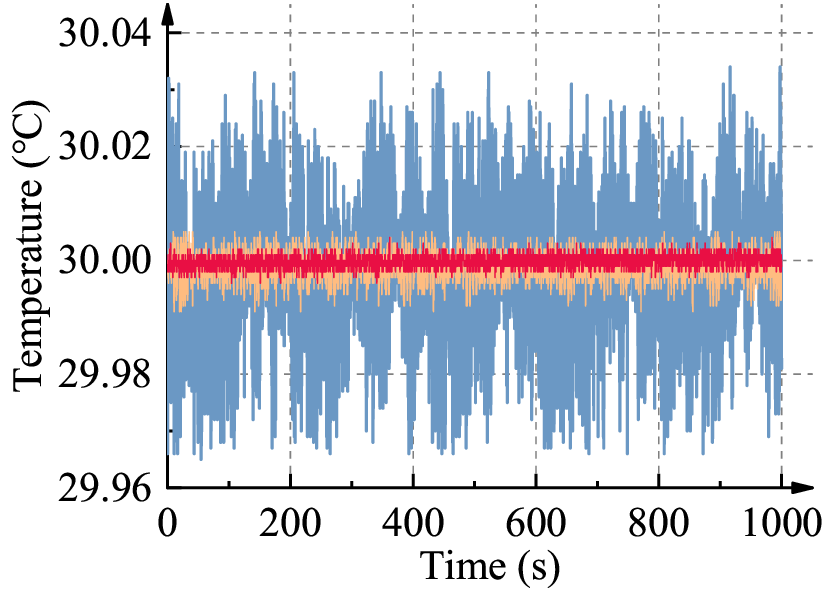}{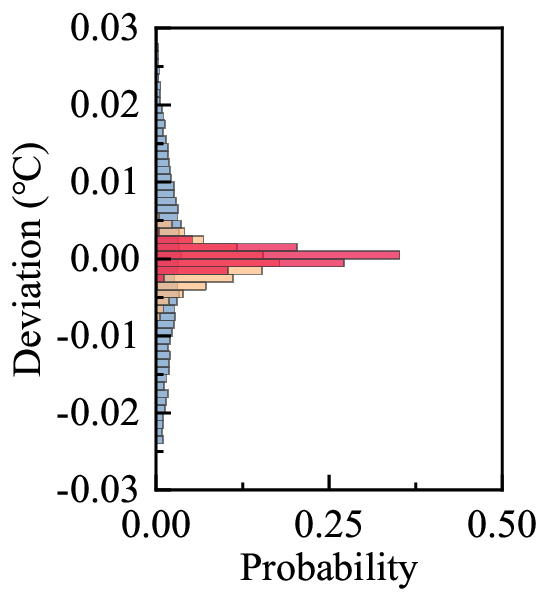}
	\\
	\vspace{-3mm}
	
	\twoplotrow{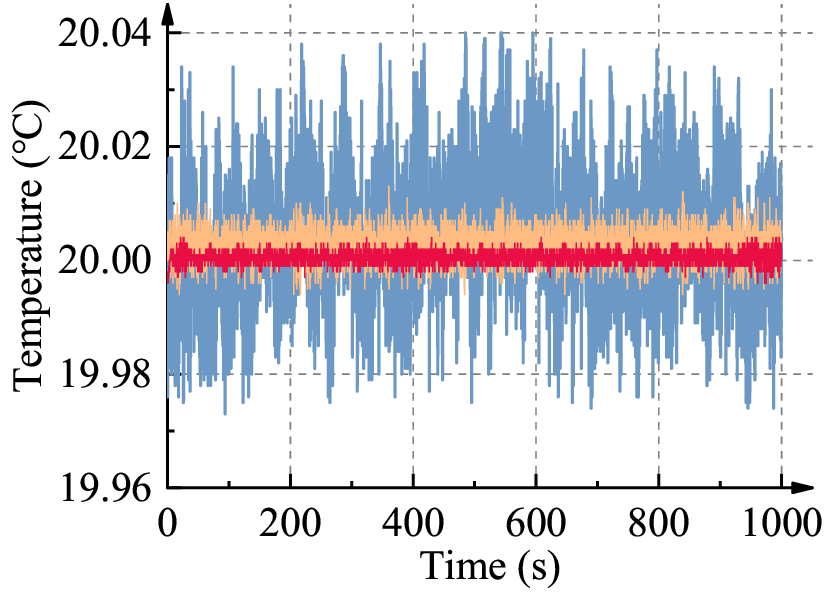}{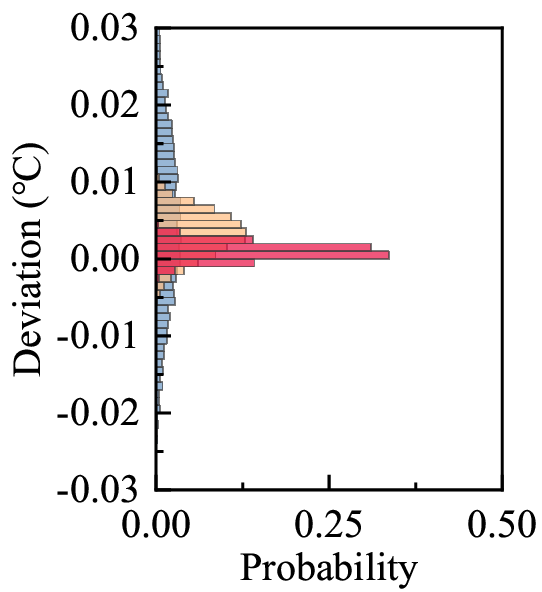}
	\twoplotrow{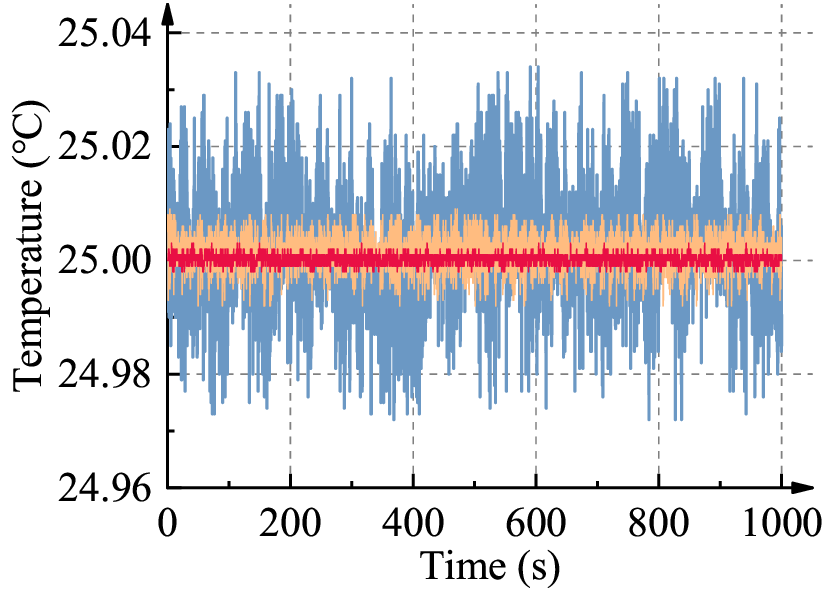}{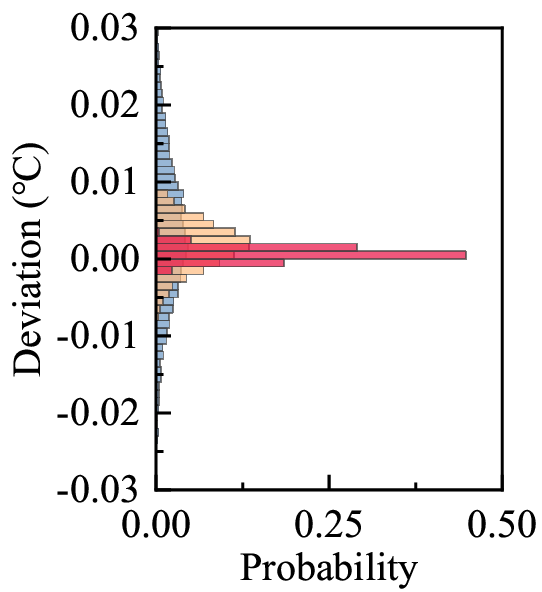}
	\twoplotrow{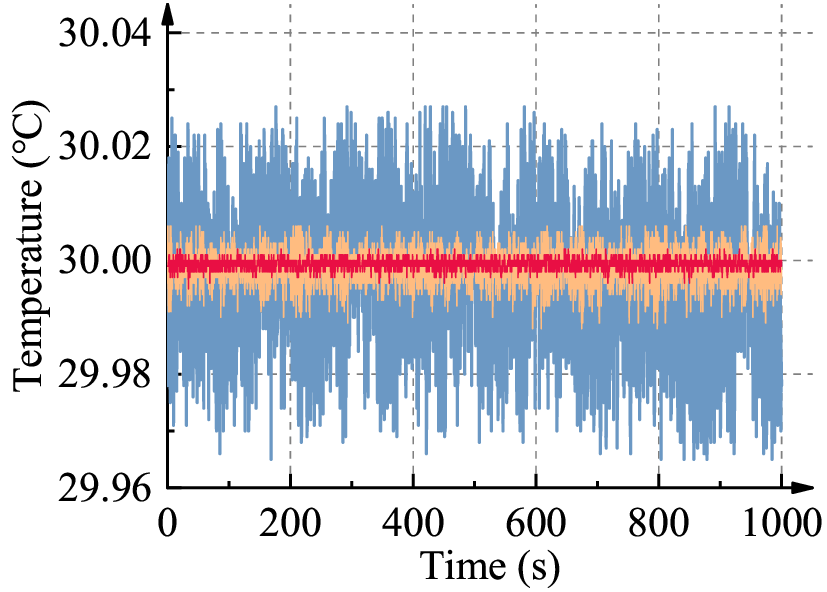}{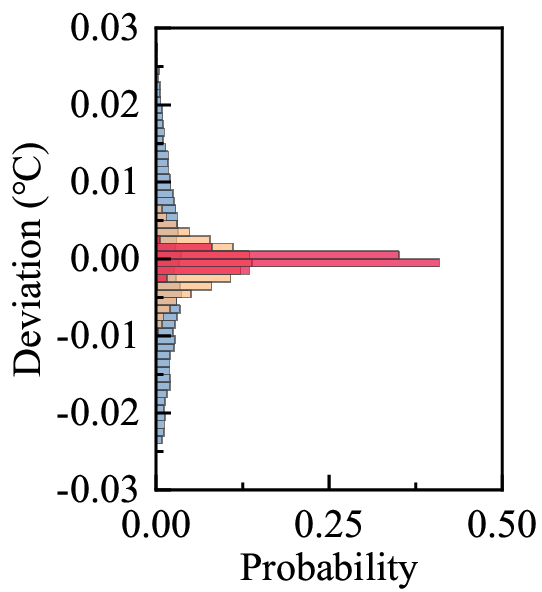}
	\vspace{-2mm}
	\caption{Temperature temporal stability and corresponding error distributions. (a) $\SI{20}{\celsius}$ at $0.5\text{ W}$. (b) $\SI{25}{\celsius}$ at $0.5\text{ W}$. (c) $\SI{30}{\celsius}$ at $0.5\text{ W}$. (d) \SI{20}{\celsius} at $1\text{ W}$. (e) \SI{25}{\celsius} at $1\text{ W}$. (f) $\SI{30}{\celsius}$ at $1\text{ W}$. (g) \SI{20}{\celsius} at $1.5\text{ W}$. (h) \SI{25}{\celsius} at $1.5\text{ W}$. (i) \SI{30}{\celsius} at $1.5\text{ W}$.}
	\label{fig7}
\end{figure}

The comparative performance of the three control algorithms across nine scenarios is presented in Fig. \ref{fig7}. Evidently, the improved efficacy of the two nonlinear MPC strategies over the linear approach validates the critical role of neural network-based predictive modeling in TA temperature control. It should be noted that the conventional PID controller performs even worse than the linear MPC, consequently excluded for conciseness. Detailed statistical metrics for each scenario are summarized in Table \ref{table3}. Both neural network-based control methods achieve significant performance improvements, with $R$ reductions exceeding $69.1\%$ and $\sigma$ reductions surpassing $62.9\%$ relative to the linear method.

\begin{table}
	\renewcommand{\arraystretch}{1.3}
	\caption{Comparison between the Proposed PI-GRU MPC Strategy and Conventional Methods}
	\centering
	\label{table4}
	\resizebox{0.8\columnwidth}{!}{
		\begin{tabular}{l l l}
			\hline\hline 
			\multicolumn{1}{c}{Method} & \multicolumn{1}{c}{Laser Power} & \multicolumn{1}{c}{\pbox{20cm}{Stability}}  \\
			\hline\\[-4mm]
			\pbox{20cm}{PID based on the heuristic Ziegler-\\Nichols method  \cite{pokryshkin2020}} & \pbox{20cm}{$10{\text{ mW}}$ \\\hphantom{1}} & \pbox{20cm}{$\SI{0.008}{\celsius} \mathord{\left/{\vphantom {{\SI{0.008}{\celsius}} 1000\text{ s}}} \right.\kern-\nulldelimiterspace} 1000\text{ s}$ @ $\SI{13}{\celsius}$ \\\hphantom{1}}\\[4mm]
			
			\pbox{20cm}{Structural optimization and PID \\algorithm  \cite{xu2020}} & \pbox{20cm}{$20{\text{ mW}}$\\\hphantom{1}} & \pbox{20cm}{$\SI{0.012}{\celsius} \mathord{\left/{\vphantom {{\SI{0.012}{\celsius}} 10000\text{ s}}} \right.\kern-\nulldelimiterspace} 10000\text{ s}$ @ $\SI{25}{\celsius}$\\\hphantom{1}}\\[4mm]
			
			\pbox{20cm}{Micro-channel structure for heat \\dissipation and active control  \cite{huang2024}}& \pbox{20cm}{$530.3{\text{ mW}}$\\\hphantom{1}} & \pbox{20cm}{$\SI{0.1}{\celsius} \mathord{\left/{\vphantom {{\SI{0.1}{\celsius}} 50\text{ s}}} \right.\kern-\nulldelimiterspace} 50\text{ s}$ @ $\SI{20.17}{\celsius}$\\\hphantom{1}}\\[4mm]
			
			\pbox{20cm}{Fuzzy PID algorithm based on \\mechanism analysis and system \\identification  \cite{zhao2022a}}& \pbox{20cm}{$10{\text{ mW}}$\\\hphantom{1}\\\hphantom{1}} & \pbox{20cm}{$\SI{0.009}{\celsius} \mathord{\left/{\vphantom {{\SI{0.009}{\celsius}} 1800\text{ s}}} \right.\kern-\nulldelimiterspace} 1800\text{ s}$ @ $\SI{20}{\celsius}$\\\hphantom{1}\\\hphantom{1}}\\[7mm]
			
			\pbox{20cm}{PI-GRU MPC\\\hphantom{1}\\\hphantom{1}} & \pbox{20cm}{$1.5{\text{ W}}$\\\hphantom{1}\\\hphantom{1}} & \pbox{20cm}{$\SI{0.008}{\celsius} \mathord{\left/{\vphantom {{\SI{0.008}{\celsius}} 1000\text{ s}}} \right.\kern-\nulldelimiterspace} 1000\text{ s}$ @ $\SI{20}{\celsius}$\\
				$\SI{0.007}{\celsius} \mathord{\left/{\vphantom {{\SI{0.007}{\celsius}} 1000\text{ s}}} \right.\kern-\nulldelimiterspace} 1000\text{ s}$ @ $\SI{25}{\celsius}$\\
				$\SI{0.010}{\celsius} \mathord{\left/{\vphantom {{\SI{0.010}{\celsius}} 1000\text{ s}}} \right.\kern-\nulldelimiterspace} 1000\text{ s}$ @ $\SI{30}{\celsius}$}\\[4mm]
			\hline\hline
		\end{tabular}
	}
\end{table}

Although the data-driven baseline and the proposed approach perform comparably at $0.5{\text{ W}}$, remarkable distinctions emerge at higher power extrapolation scenarios. PI-GRU MPC achieves stronger robustness compared to GRU MPC, reducing $R$ by over $57.9\%$ and $\sigma$ by over $65.6\%$ at power levels of $1.5{\text{ W}}$. Particularly, PI-GRU MPC maintains Allan deviations $\sigma \left( {100{\text{ s}}} \right)$ at $1.5{\text{ W}}$ comparable to those achieved at  $0.5{\text{ W}}$, thereby achieving demanding long-term stability paramount to biomagnetic imaging. In summary, PI-GRU MPC provides outperforming temperature control stability and exceptional power-level generalization, ideal for ultra-stable TA lasers.

To underscore the key innovation and outstanding performance of our approach, we undertake a comprehensive comparison between the proposed strategy and conventional semiconductor laser temperature stabilization methods reported in prior studies, as illustrated in Table \ref{table4}. The proposed method not only proves stabilization performance on par with control techniques tailored for low-power lasers but also offers significant applicability to high-power regimes. Superior to conventional methods, PI-GRU MPC enables adaptable deployment of TA lasers in quantum sensing systems demanding diverse operational scenarios.

\section{Conclusion}\label{sec5}
In conclusion, this study proposes an innovative framework integrating an encoder-decoder PI-GRU architecture with MPC strategy, effectively addressing temperature control challenges in nonlinear TA laser systems. By embedding soft physical constraints derived from a  lumped-parameter thermal model, a high-precision cross-power temperature predictive model is trained exclusively on low-power samples. Moreover, a hierarchical parallel architecture is established to execute multi-stage computational tasks through tiered batch processing, guaranteeing real-time solutions in closed-loop control. Experimental validations demonstrate its superb temperature stability in multi-power laser scenarios, where the end-to-end deep learning architecture overcomes traditional models' limitations in nonlinear system modeling. Although further improvements in transient behavior remain feasible, the PINN-enhanced MPC strategy presents a novel paradigm for intelligent control of complex coupled systems.

%
%
%






\bibliographystyle{elsarticle-num}
\bibliography{BIB_TA}

%
%
%
\end{document}